\UseRawInputEncoding
\documentclass[preprintnumbers,a4paper,11pt]{article}
\usepackage{authblk,abstract}
 
\usepackage{caption,subcaption}
\usepackage{graphicx} 
\usepackage{rotating,float,afterpage}
\usepackage{calrsfs}
\usepackage{slashed,cancel}
\usepackage{relsize}
\RequirePackage[T1]{fontenc}
\usepackage[utf8x]{inputenc}
\usepackage{amsmath,graphicx,epsfig,amssymb,multirow}
\RequirePackage{mathptmx,flushend}      
\usepackage[12hr]{datetime} 
\RequirePackage[numbers,sort&compress]{natbib}
\usepackage{xcolor}
\usepackage[hidelinks]{hyperref}
\hypersetup{
	colorlinks,
	linkcolor={blue!80!black},
	citecolor={blue!80!black},
	urlcolor={blue!80!black}
}

\usepackage[a4paper, inner=2.2cm, outer=2.2cm, top=2.2cm,bottom=2.5 cm, binding offset=0 cm]{geometry}
\usepackage{color}
\definecolor{myred}{rgb}{0.6,0,0} 
\definecolor{myblue}{rgb}{0,0.2,0.4}
\definecolor{mygreen}{rgb}{0,0.9,0.1}
\definecolor{hc}{rgb}{.9,0.1,0.7}
\definecolor{hcout}{rgb}{.9,0.7,0.9}
\definecolor{Orange}{rgb}{1.,0.65,0.}

\allowdisplaybreaks

\def\wtil#1{\widetilde{#1}}
\def\MGvATNLO{{\tt {\sc MadGraph5}\_aMC@NLO}}
\def\l{\left}
\def\r{\right}

		\title{Breaking down the entire spectrum of spin correlations  of a pair of particles   involving fermions and gauge bosons}
		\author[a]{Rafiqul Rahaman\thanks{rafiqulrahaman@hri.res.in}}
		\affil[a]{Regional Centre for Accelerator-based Particle Physics, Harish-Chandra Research Institute, A CI of Homi Bhabha National Institute, 
			Chhatnag Road, Jhunsi, Prayagraj 211019, India}
		\author[b]{Ritesh K. Singh\thanks{ritesh.singh@iiserkol.ac.in}}
		\affil[b]{Department of Physical Sciences, Indian Institute of Science
			Education and Research Kolkata, Mohanpur, 741246, India}

		\date{HRI-RECAPP-2021-010}

\begin{document}

\maketitle
	\begin{abstract}
	We discuss a formalism for the spin correlations and polarizations in two-particle systems with spins half-half, half-one and one-one, and provide the connections between the polarizations and correlations with the joint angular distributions of decay products  identifying the asymmetries for them. We demonstrate the formalism in  the  processes $e^-e^+\to t\bar{t}$,  $gb \to tW^-$  and $e^-e^+\to ZZ$ in the standard model as examples.  We investigate the effect of some anomalous couplings on the polarizations and spin correlations  in the  processes $e^-e^+\to t\bar{t}$,  $gb \to tW^-$ and $u\bar{d}\to ZW^+$ at parton level and compare their strengths. The spin correlations have the potential to provide a significant improvement over the polarizations in probing the anomalous couplings.
\end{abstract}

	\section{Introduction}
	The scaler sector of the standard model (SM) particle spectrum will be affected by possible new physics  beyond the SM (BSM) needed to address phenomena, such as dark matter, neutrino oscillation, baryogenesis etc. These  BSM physics will lead to a modified electroweak sector with modified interactions among the Higgs bosons, top quark, and the gauge bosons. Precise measurement of the interactions in terms of strength and tensorial structure,  thus, may uncover new physics at colliders. 
	
	Polarizations of top quark and massive gauge bosons are interesting tools for such precision measurement, as they are sensitive to the modification of the interactions involved. The top quark, $Z$ and $W$ bosons being massive, they decay immediately after they are produced, letting their decay products carry their polarization as well as spin correlation information. In general a spin-$s$ particle contains $(2s+1)^2-1=4s(s+1)$ polarization parameters, e.g., a spin-$1/2$ particle has $3$ vector polarizations,  and a spin-$1$ particle carries $5$ tensor polarizations along with $3$ vector polarizations~\cite{Bourrely:1980mr,Boudjema:2009fz}. These polarization parameters can
	be calculated from the production process as well as from the angular distributions of the decay products~\cite{Boudjema:2009fz,Rahaman:2016pqj}. Like the polarizations, a system of two particles $A$ and $B$ with spins $s_A$ and $s_B$ contains $4s_A(s_A+1)\times 4s_B(s_B+1)$
	spin correlation parameters along with  $4s_A(s_A+1)+4s_B(s_B+1)$ polarization parameters. These spin correlation parameters can also be calculated from the production
	process of the two-particle system as well as from the joint angular distributions of their decay products. 
	
	There have been a lot of interest in top quark polarizations~\cite{Kane:1991bg,Jezabek:1994zv,Hikasa:1999wy,Godbole:2006tq,Perelstein:2008zt,Huitu:2010ad,Choudhury:2010cd,Arai:2010ci,Gopalakrishna:2010xm,Godbole:2010kr,Godbole:2011vw,Krohn:2011tw,Rindani:2011pk,Cao:2011hr,Rindani:2011gt,Bhattacherjee:2012ir,Fajfer:2012si,Biswal:2012dr,Belanger:2013gha,Baumgart:2013yra,Godbole:2015bda,Rindani:2015vya,Rindani:2015dom,Behera:2018ryv,Arhrib:2018bxc,Wu:2018xiz,Zhou:2019alr,Arhrib:2019tkr,Patrick:2019nhv}  along with spin correlations of $t\bar{t}$ system~\cite{Cheung:1996kc,Arai:2004yd,Arai:2007ts,Arai:2009cp,Yue:2010hy,Degrande:2010kt,Cao:2010nw,Baumgart:2011wk,Barger:2011pu,Fajfer:2012si,Kiers:2014uqa,Bernreuther:2013aga,Bernreuther:2015yna,Aguilar-Saavedra:2018ggp,Ravina:2021kpr} in probing new  physics.
	The top quark polarizations and spin correlations have also been measured at the Tevatron~\cite{Aaltonen:2010nz,Abazov:2011ka,Abazov:2011qu,Abazov:2011gi,Abazov:2012oxa,Abazov:2015psg,Lee:2018lgy} as well as at the large hadron collider ( LHC)~\cite{ATLAS:2012ao,Aad:2013ksa,Aad:2014pwa,Aad:2014mfk,Khachatryan:2015tzo,Aad:2015bfa,Tiko:2016brs,Khachatryan:2016xws,Aaboud:2016bit,CMS:2018jcg,Aaboud:2019hwz}. 
	Lately,  polarization for $Z$ and $W$ bosons also got some attention in probing new physics~\cite{Abbiendi:2000ei,Rahaman:2016pqj,Rahaman:2017qql,Rahaman:2017aab,Nakamura:2017ihk,Aguilar-Saavedra:2017zkn,Rahaman:2018ujg,Rao:2018abz,Renard:2018tae,Renard:2018bsp,Renard:2018lqv,Rahaman:2019mnz,Rahaman:2019lab}. The predictions of $Z$ and $W$ polarizations are also being developed within the SM  including next-to-leading order effects~\cite{Baglio:2018rcu,Baglio:2019nmc,Denner:2020bcz,Denner:2020eck,Denner:2021csi,Poncelet:2021jmj}. The polarizations of $Z$ and $W$ have also been measured recently at the LHC in $ZW$ production process~\cite{Aaboud:2019gxl,CMS:2021lix}. The decay  angular correlations, in case of vector boson pair production,  are discussed before in Refs.~\cite{Hagiwara:1986vm,Ohnemus:1994ff,Dixon:1998py,Dixon:1999di}. 
	For the spin correlation, although the $t\bar{t}$, i.e., half-half spin system, correlation  have been talked about before, only a subset of the full nine correlations are discussed. 
	Furthermore,  although spin-correlated spin density matrix (SDM) and decay distribution with correlations have been discussed earlier for the $t\bar{t}$ pair production~\cite{Bernreuther:1993df,Bernreuther:1993hq,Mahlon:1995zn,Brandenburg:1996df,Bernreuther:2000yn,Mahlon:2010gw}, no clear connection between them has been made before. 
	For the $tV$ ($V=Z/W$) or $VV$ pair production, i.e., half-one or one-one spin systems, there exist no formalism for the spin correlated production matrix or SDM, and thus no existence of a connection between the decay distributions and spin correlations.
	In this article, we present a formalism for the spin correlated density matrix and the connection between joint  decay angular distributions and the full-rank spin correlations in three spin systems, namely half-half, half-one, and one-one. 
	 The presented formalism is validated in some simple SM  processes comparing the analytically obtained values for the spin correlations to those numerically extracted values from the angular distributions of the decay products obtained in  Monte-Carlo simulation. The potential of the spin correlations is investigated in probing anomalous couplings, parameterized by effective operators, in some simple partonic processes without realistic effects to make an ansatz for real scenarios with collider data.

	We proceed with a general description of production and decay of two spin full particles in section~\ref{sec:spin-formalism} followed by  discussing the individual polarizations of spin-$1/2$ and spin-$1$ particles in section~\ref{sec:polarization-single}. Next, we present the formalism for spin correlation for the three spin systems, i.e., half-half, half-one, and one-one successively in section~\ref{sec:spin-corr}. In section~\ref{sec:sm-example}, we demonstrate the formalism in the SM by showing the agreement between the analytically calculated  correlations from the production process and the numerically calculated values from \MGvATNLO~\cite{Alwall:2014hca} simulations using decay distributions in  three processes $e^-e^+\to t\bar{t}$, $gb \to tW^-$ and $e^-e^+\to ZZ$ as a proof of principle. In section~\ref{sec:bsm-examples}, we discuss how spin correlations can help in probing new physics in parallel to polarization in three processes $e^-e^+\to t\bar{t}$,  $gb \to tW^-$ and $u\bar{d}\to ZW^+$ at parton level and without initial state of parton distribution functions (PDFs) folding, for simplicity. We summarize in section~\ref{sec:discussion}.

\section{Formalism}\label{sec:spin-formalism}
\begin{figure}
\begin{center}
\includegraphics[width=0.6\textwidth]{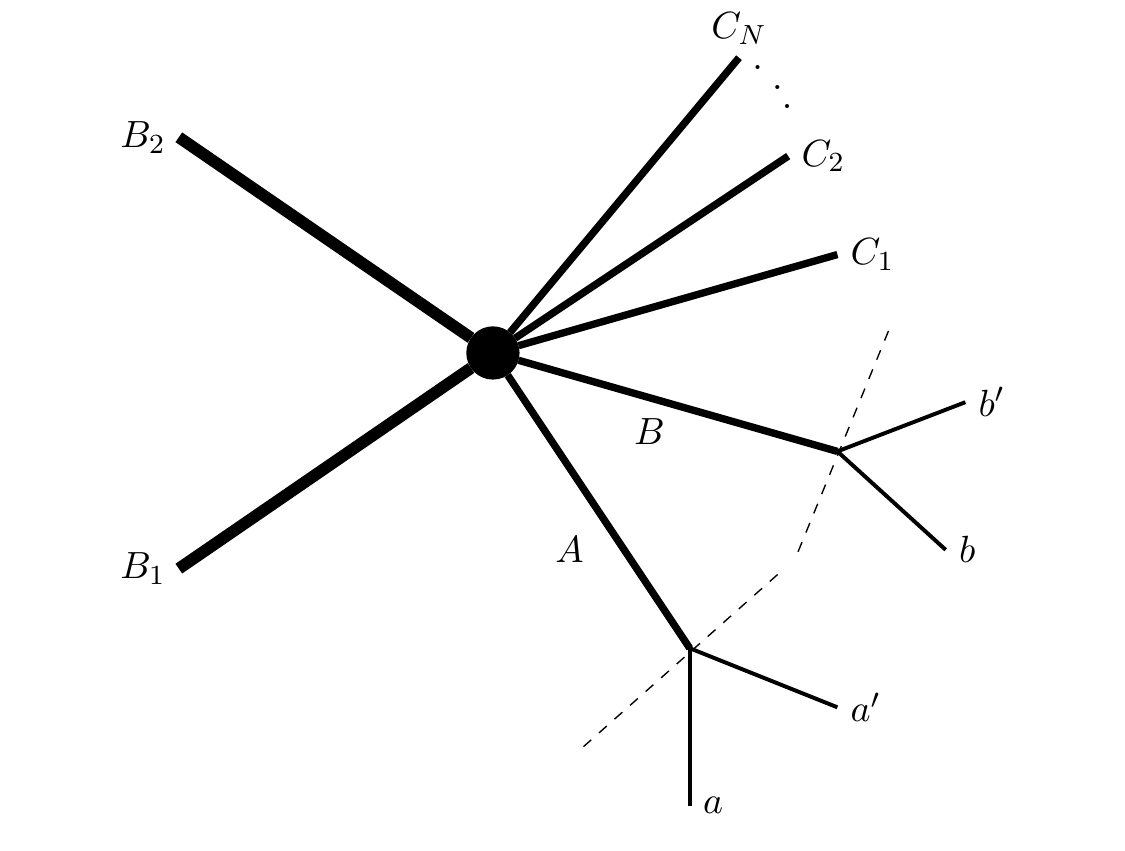}
\end{center}
		\caption{\label{fig:AB-production-decay} Schematic diagram for the production of two particles $A$ and $B$ along with bunch of other particles $C_1,~C_2,\ldots,~C_N$ by the collision of beam particles $B_1$ and $B_2$. The particles $A$ and $B$ decay subsequently as  $A\to aa^\prime$ and $B\to b b^\prime$. The dashed lines are shown to separate the production part from the decay parts.} 
\end{figure}
	To describe the spin correlations along with spin-polarizations of a system of two spin-full particles, let us consider the production and decay of two massive spin-full unstable particles $A$ and $B$ in a general process  $B_1B_2\to A B C_1C_2\ldots C_N$ with  $A\to aa^\prime$ and $B\to b b^\prime$, as shown in Fig.~\ref{fig:AB-production-decay}.  The differential rate for such a process can be expressed as~\cite{Boudjema:2009fz}, 
	\begin{eqnarray}\label{eq:whole-process}
		d\sigma = \mathlarger{\mathlarger{‎‎\sum}}_{\lambda_A,\lambda_A^\prime,\lambda_B,\lambda_B^\prime}&&\left[ \frac{1}{I_{B_1B_2}}  \rho_{AB}^\prime(\lambda_A,\lambda_A^\prime,\lambda_B,\lambda_B^\prime) (2\pi)^4\delta^4\l(p_{B_1}+p_{B_2}-p_A-p_B-\sum_{i=1}^{N}p_{C_i}\r)\right.\nonumber\\ &\times&\left.\l(\dfrac{d^3p_A}{(2\pi)^3 2E_A}\r)\l( \dfrac{d^3p_B}{(2\pi)^3 2E_B}  \r)    \prod_{i=1}^{N}\l( \dfrac{d^3p_{C_i}}{(2\pi)^3 2E_{C_i}}\r)    \right]
		\nonumber\\
		&\times& \left[\frac{1}{\Gamma_A}\frac{(2\pi)^4}{2m_A} \Gamma_A^\prime(\lambda_A,\lambda_A^\prime)\delta^4\l(p_A-p_{a}-p_{a^\prime} \r) \l(\dfrac{d^3p_{a}}{(2\pi)^3 2E_{a}}\r) \l(\dfrac{d^3p_{a^\prime}}{(2\pi)^3 2E_{a^\prime}}\r)    \right] \nonumber\\
		&\times& \left[\frac{1}{\Gamma_B}\frac{(2\pi)^4}{2m_B} \Gamma_B^\prime(\lambda_B,\lambda_B^\prime)\delta^4\l(p_B-p_{b}-p_{b^\prime} \r) \l(\dfrac{d^3p_{b}}{(2\pi)^3 2E_{b}}\r) \l(\dfrac{d^3p_{b^\prime}}{(2\pi)^3 2E_{b^\prime}}\r)    \right]\nonumber\\
	\end{eqnarray} 
	in the narrow-width approximation of the unstable particles $A$ and $B$, allowing the factorization of production part (in the first square bracket) and the decay parts (in the second and third square brackets). 
	Here, $p_X$, $m_X$ and $E_X$ with $X\in \{B_i,A,B,C_i,a,a^\prime,b,b^\prime\}$ are the four-momenta, mass and energy of the particles, respectively; 
	the flux factor $I_{B_1B_2}$ is given by $I_{B_1B_2}=4\sqrt{\l(p_{B_1}\cdot p_{B_2}\r)^2-m_{B_1}^2 m_{B_2}^2}$ with
	$m_{B_i}$  the  mass of the particles $B_i$;   $\Gamma_{A/B}$ are the total decay width of $A/B$. 
	The $\lambda_{A/B}$,  $\lambda_{A/B}^\prime$ are the helicities of $A/B$ and they can take values in the range of 
	$$\lambda_{A/B},\lambda_{A/B}^\prime \in\l[-s_{A/B},-s_{A/B}+1,\ldots,s_{A/B}\r] $$
	with $s_{A}$ and $s_B$ as  the spins of the particle $A$ and $B$, respectively.
	The helicities of all particles other than $A$ and $B$ are suppressed, i.e., helicities are summed over the $C_i$ and averaged over the $B_i$\footnote{For polarized beams, one has to use the initial state polarization density matrix in  Eq.~(\ref{eq:whole-process}), e.g., see Ref.~\cite{Rahaman:2017qql}.}.
	The phase space integration of the decay products of $A$ and $B$ can be performed   in the rest of $A$ and $B$ separately without any loss of generality. 
	The production part or the production density matrix of $A$ and $B$
	can be expressed as,
	\begin{eqnarray}\label{eq:production_density_matrix}
		\rho_{AB}(\lambda_A,\lambda_A^\prime,\lambda_B,\lambda_B^\prime) &= & \frac{1}{I_{B_1B_2}} \int \rho_{AB}^\prime(\lambda_A,\lambda_A^\prime,\lambda_B,\lambda_B^\prime) (2\pi)^4\delta^4\l(p_{B_1}+p_{B_1}-p_A-p_B-\sum_{i=1}^{N}p_{C_i}\r)\nonumber\\ &\times&\l(\dfrac{d^3p_A}{(2\pi)^3 2E_A}\r)\l( \dfrac{d^3p_B}{(2\pi)^3 2E_B}\r)      \prod_{i=1}^{N} \l(\dfrac{d^3p_{C_i}}{(2\pi)^3 2E_{C_i}} \r)   
	\end{eqnarray} 
	with $$\rho_{AB}^\prime(\lambda_A,\lambda_A^\prime,\lambda_B,\lambda_B^\prime) = {\cal M} (\lambda_A,\lambda_B)\times {\cal M}^\dagger(\lambda_A^\prime,\lambda_B^\prime),$$ ${\cal M}(\lambda_A,\lambda_B)$ being the helicity matrix amplitude with helicities $\lambda_A$ and $\lambda_B$.
	The  cross section for the production  of $A$ and $B$  would be,  
	\begin{equation}{\label{eq:sigma-AB}}
		\sigma_p = \text{Tr}\l[\rho_{AB}(\lambda_A,\lambda_A^\prime,\lambda_B,\lambda_B^\prime)  \r]	
		= \sum_{\lambda_A,\lambda_B} \rho_{AB}(\lambda_A,\lambda_A,\lambda_B,\lambda_B). 
	\end{equation}
	Here, only the diagonal elements of the $(2s_A+1)(2s_B+1) \times  (2s_A+1)(2s_B+1)$ dimensional production density matrix $\rho_{AB}$ enter, while  all  the elements in a certain combination contain polarization and spin correlation information of $A$ and $B$. Thus the normalized production density matrix can be compared to the spin density matrix of the system $A$ and $B$. We rewrite, 
	\begin{equation}\label{eq:def-PAB}
		\rho_{AB}(\lambda_A,\lambda_A^\prime,\lambda_B,\lambda_B^\prime)=\sigma_p \times P_{AB(2s_A,2s_B)}(\lambda_A,\lambda_A^\prime,\lambda_B,\lambda_B^\prime)
	\end{equation}
	with $P_{AB(2s_A,2s_B)}(\lambda_A,\lambda_A^\prime,\lambda_B,\lambda_B^\prime)$ as the normalized production density matrix,
	we call it polarization-correlation density matrix.
	The decay part of $A$, after partial phase-space integration, can be expressed as, 
	\begin{eqnarray}\label{eq:general-decay-part}
		&&\mathlarger{\mathlarger{\int}}\frac{1}{\Gamma_A}\frac{(2\pi)^4}{2m_A} \Gamma_A^\prime(\lambda_A,\lambda_A^\prime)\delta^4\l(p_{A}-p_{a}-p_{a^\prime} \r) \l(\dfrac{d^3p_{a}}{(2\pi)^3 2E_{a}} \r)\l(\dfrac{d^3p_{a^\prime}}{(2\pi)^3 2E_{a^\prime}} \r)
		\nonumber \\
		&=& \frac{Br\left(A\to a a^\prime \right)(2s_{A}+1)}{4\pi}\Gamma_{A(2s_A)}(\lambda_{A},\lambda_{A}^\prime) d\Omega_{a/a^\prime},
	\end{eqnarray}
	where $Br\left(A\to a a^\prime \right)$ is the branching fraction of $A$ decaying to $a a^\prime$.  The quantity $\Gamma_{A}(\lambda_{A},\lambda_{A}^\prime)$ is the  decay density matrix normalized to unit trace, and 
	$d\Omega_{a}=\sin\theta_{a} d\theta_{a} d\phi_{a}$ is the measure of  solid angle of the daughter 
	$a$. The decay part of $B$ can also be expressed in the same way as done for $A$.
	Combining the production density matrix of $A$ and $B$ in Eq.~(\ref{eq:production_density_matrix}) and their decay density matrices in Eq.~(\ref{eq:general-decay-part}) according to Eq.~(\ref{eq:whole-process}), one obtains the normalized joint angular distribution of the decay products as, 
	\begin{eqnarray}\label{eq:norm_dist}
		\dfrac{1}{\sigma}\dfrac{d^2\sigma}{d\Omega_ad\Omega_b}=\frac{2s_A+1}{4\pi}\frac{2s_B+1}{4\pi}
		\mathlarger{\mathlarger{‎‎\sum}}_{\lambda_A,\lambda_A^\prime,\lambda_B,\lambda_B^\prime}^{}  &&P_{AB(2s_A,2s_B)}(\lambda_A,\lambda_A^\prime,\lambda_B,\lambda_B^\prime) \nonumber\\
		&&\times\Gamma_{A(2s_A)}(\lambda_A,\lambda_A^\prime)
		\times\Gamma_{B(2s_B)}(\lambda_B,\lambda_B^\prime) , 
	\end{eqnarray}
	where $\sigma= 	\sigma_p  \times Br\left(A\to a a^\prime \right)\times Br\left(B\to b b^\prime \right)$ is the total cross section of the production of $A$ and $B$ followed by  their decays.

	The polarizations of $A$ and $B$ and their spin correlations,  embedded into their polarization-correlations density matrix $P_{AB(2s_A,2s_B)}$, are transferred
	to the distributions of  their decay products $a$ and $b$. One can obtain the polarizations and spin correlations from the production part as well as form the joint angular distributions of the decay products. In this work, we  consider a system with a pair of particles  combining only with   spin-$1/2$ and spin-$1$ particles, i.e., half-half, half-one, and one-one spin systems.  In the next sections, we describe how the 
	spin correlations along with the polarizations of two particles in these systems can be computed from the production part as well as from the decay distributions.  First, we briefly present the scenarios with a single spin-$1/2$ and spin-$1$ particles in the next section for completeness.   
	
	\subsection{Polarizations of a single particle}\label{sec:polarization-single}
	The polarization density matrix or the spin-density matrix (SDM) of a spin-$s$ particle can be expressed  with the irreducible spin tensors up to rank $2s$, i.e., identity matrix, linear ( for spin-$1/2$ and spin-$1$), bilinear (only for spin-$1$)  combinations of standard spin matrices. The SDM can be represented  in terms of multi-pole parameters or can be given in Cartesian form~\cite{Bourrely:1980mr}. The properties of the density matrix will then be specified by the expansion coefficients. In Cartesian form,  the polarization density matrix of spin-$1/2$ and 
	spin-$1$ particles can be expressed as~\cite{Bourrely:1980mr,Boudjema:2009fz},
	\begin{eqnarray}\label{eq:pol-density-half}
		P_{f(1)}(\lambda,\lambda^\prime) &=& \frac{1}{2}\Big[ \mathbb{I}_{2\times 2} + \vec{p}\cdot\vec{\tau}\Big],~~\lambda,\lambda^\prime\in[+1,-1],\nonumber\\
		&=&\frac{1}{2}\left[
		\begin{tabular}{cc}
			$1+p_z$&$p_x - i p_y$\\
			$p_x + i p_y$&$1-p_z$
		\end{tabular}
		\right]~~~~~~~\text{and}
	\end{eqnarray}
	\begin{eqnarray}\label{eq:pol-desnity-one}
		P_{V(2)}(\lambda,\lambda^\prime)&=&\dfrac{1}{3}\l[\mathbb{I}_{3\times 3} +\dfrac{3}{2} \vec{p}.\vec{S}
		+\sqrt{\dfrac{3}{2}} T_{ij}\big(S_iS_j+S_jS_i\big) \r],~~\lambda,\lambda^\prime\in[+1,0,-1],\nonumber\\
		&=&
		\renewcommand{\arraystretch}{1.5}
		\left[
		\begin{tabular}{lll}
			$\frac{1}{3}+\frac{p_z}{2}+\frac{T_{zz}}{\sqrt{6}}$ &
			$\frac{p_x -ip_y}{2\sqrt{2}}+\frac{T_{xz}-iT_{yz}}{\sqrt{3}}$ &
			$\frac{T_{xx}-T_{yy}-2iT_{xy}}{\sqrt{6}}$ \\
			$\frac{p_x +ip_y}{2\sqrt{2}}+\frac{T_{xz}+iT_{yz}}{\sqrt{3}}$ &
			$\frac{1}{3}-\frac{2 T_{zz}}{\sqrt{6}}$ &
			$\frac{p_x -ip_y}{2\sqrt{2}}-\frac{T_{xz}-iT_{yz}}{\sqrt{3}}$ \\
			$\frac{T_{xx}-T_{yy}+2iT_{xy}}{\sqrt{6}}$ &
			$\frac{p_x +ip_y}{2\sqrt{2}}-\frac{T_{xz}+iT_{yz}}{\sqrt{3}}$ &
			$\frac{1}{3}-\frac{p_z}{2}+\frac{T_{zz}}{\sqrt{6}}$
		\end{tabular}\right],
	\end{eqnarray}
	respectively. Repeated indices are  summed over.
	Here, $\tau_i$ and $S_i$ are the spin matrices in spin-$1/2$ and spin-$1$ basis (given in appendix~\ref{sec:spin-matrices}), respectively.
	The $p_i$s, components of  a three vector $\vec{p}=\{p_x,p_y,p_z\}$, are vector polarizations of the particles;  $T_{ij}$, components  of a second-rank symmetric trace-less tensor, are tensor polarizations of the spin-$1$ particle. For a spin-$1/2$ particle, there are three vector polarizations, while for a spin-$1$ particle, there are three vector and five independent tensor polarizations. These polarization parameters can be estimated by comparing the  polarization density matrices to the respective normalized production density matrices. 
	Combining the polarization density matrices with their respective normalized decay density matrices obtained at their rest frame (given in appendix~\ref{sec:spin-matrices} in Eqs. ~(\ref{eq:decay-half}) and (\ref{eq:decay-one}) ), one  obtains the angular distributions of decay products for  $f(1)$  and $V(2)$ as~\cite{Boudjema:2009fz}, 
	\begin{equation}\label{eq:norm_dist-half}
		\frac{1}{\sigma_f}\frac{d\sigma_f}{d\Omega}=\frac{1}{4\pi}\Big[   
		1+\alpha p_x \sin\theta\cos\phi +\alpha p_y\sin\theta\sin\phi +\alpha p_z\cos\theta
		\Big]~~~\text{and}
	\end{equation}
	\begin{eqnarray}\label{eq:norm_dist-one}
		\frac{1}{\sigma_V}  \frac{d\sigma_V}{d\Omega} &=&\frac{3}{8\pi} \left[
		\left(\frac{2}{3}-(1-3\delta) \ \frac{T_{zz}}{\sqrt{6}}\right) + \alpha \ p_z
		\cos\theta 
		+ \sqrt{\frac{3}{2}}(1-3\delta) \ T_{zz} \cos^2\theta
		\right.\nonumber\\
		&+&\left(\alpha \ p_x + 2\sqrt{\frac{2}{3}} (1-3\delta)
		\ T_{xz} \cos\theta\right) \sin\theta \ \cos\phi \nonumber\\
		&+&\left(\alpha \ p_y + 2\sqrt{\frac{2}{3}} (1-3\delta)
		\ T_{yz} \cos\theta\right) \sin\theta \ \sin\phi \nonumber\\
		&+&\left.(1-3\delta) \left(\frac{T_{xx}-T_{yy}}{\sqrt{6}} \right) \sin^2\theta
		\cos(2\phi)
		+\sqrt{\frac{2}{3}}(1-3\delta) \ T_{xy} \ \sin^2\theta \
		\sin(2\phi) \right] , 
	\end{eqnarray}
	respectively. Here, $\theta$ and $\phi$ are the polar and azimuthal angles of the decay products in the rest frame of the mother particles with their would-be momentum along $z$-direction. 
	One can construct several asymmetries related to the polarization parameters by partially integrating the angular distributions with respect to $\theta$ and $\phi$. The complete list of polarization asymmetries is given in Refs.~\cite{Rahaman:2016pqj,Rahaman:2020jll}. 
	We will list them later while discussing correlations of a pair of particles in the next section.

	\subsection{Spin correlations of a pair of particles}\label{sec:spin-corr}
	In the spin correlated density matrix of two spin-full particles, there are 
	correlations between two vector polarizations (vector-vector correlation),  correlation between a vector and a tensor polarization  (vector-tensor polarization) and  correlation between two  tensor polarizations (tensor-tensor correlation) apart from  the vector polarizations and tensor polarizations. We proceed with, first, discussing the simple spin-$1/2$ -- spin-$1/2$ correlations and then spin-$1/2$ -- spin-$1$ and spin-$1$ -- spin-$1$ subsequently.

	\subsubsection{Spin-$1/2$ -- spin-$1/2$ correlations}
	The spin correlations for a pair of spin-$1/2$ particles can be accommodated in the polarization-correlation density matrix with an outer product of two Pauli spin-$1/2$ matrices ($\tau\otimes\tau$), one for each particle. Thus the fully spin correlated polarization density matrix of a pair of spin-$1/2$ particles will be given by~\cite{Bernreuther:1993df,Bernreuther:1993hq}
	\begin{eqnarray}\label{eq:spin-density-half-half}
		P_{AB(1,1)}\l(\lambda_A,\lambda_A^\prime,\lambda_B,\lambda_B^\prime\r) = \frac{1}{\l(2\times\frac{1}{2}+1\r)^2}\Biggr[ \underbrace{\mathbb{I}_{2\times 2}\otimes \mathbb{I}_{2\times 2}}_{\mathbb{I}_{4\times 4}} + \vec{p}^A\cdot\vec{\tau}\otimes \mathbb{I}_{2\times 2} + \mathbb{I}_{2\times 2} \otimes \vec{p}^B\cdot\vec{\tau} \nonumber\\
		+ pp_{ij}^{AB}\tau_i\otimes\tau_j  \Biggr],~\l( i,j\in[ x\equiv 1,y\equiv 2,z\equiv 3] \r),
	\end{eqnarray}
	where $\vec{p}^{A/B}$ are the vector polarizations of $A/B$; $pp_{ij}^{AB}$, components of a second-rank tensor $pp^{AB}$, are the vector-vector correlations of $A$ and $B$ \footnote{Equation~(\ref{eq:spin-density-half-half})  can  be extended for three spin-half particles  introducing  a rank-three correlation parameter, let's say $ppp^{ABC}$, $C$ being the third spin-half particle, with a suitable outer product of the spin matrices ($\tau$).}.
	The elements of $pp^{AB}$ are completely independent of each other. However, the $pp^{AB}$ tensor can be symmetric if $A=B$, i.e., they are identical. 
	Thus, there exists a total of $3+3+9=15$ polarization and correlation parameters (see Table~\ref{tab:num-param-half-half})  in a spin-$1/2$ -- spin-$1/2$ pair of particles production. 
	\begin{table}[h!]
		\centering
		\begin{tabular}{|c|ccc|c|}\hline
			Parameter &	$p_{i}^A$ & $p_{i}^B$ & $pp_{ij}^{AB}$  & Total \\ \hline
			Number of parameter &	$3$ & $3$ &$9$ & $=15$\\ \hline
		\end{tabular}	
		\caption{\label{tab:num-param-half-half} Number of polarization and spin correlation parameters in a spin-$1/2$ -- spin-$1/2$ pair of particles production.}
	\end{table}
	The polarization ($p_i^A$, $p_i^B$) of $A$ and $B$  can be obtained either from their individual production density matrices or from the combined production density matrix of $A$ and $B$, while the vector-vector spin correlations $pp_{ij}^{AB}$ can only be obtained from their combined production density matrix.
		The expanded form of the spin correlated polarization density matrix in Eq.~(\ref{eq:spin-density-half-half})
		is given in appendix~\ref{app:corr-from-production}. The polarization and spin correlation parameters are also given  in terms of the production density matrix elements in appendix~\ref{app:corr-from-production}. 
	
	After combining the normalized spin-$1/2$ decay density matrices (Eq.~(\ref{eq:decay-half})) with the polarization-correlation density matrix $P_{AB\left(\frac{1}{2},\frac{1}{2}\right)}$ (Eq.~(\ref{eq:spin-density-half-half})) of $A$ and $B$ in accordance with the Eq.~(\ref{eq:norm_dist}), we obtain the joint angular distribution of the decay products as,
	\begin{eqnarray}\label{eq:norm_dist-half-half}
		\dfrac{1}{\sigma}\dfrac{d^2\sigma}{d\Omega_{a}d\Omega_{b}}&=&\frac{2}{4\pi}\frac{2}{4\pi}
		\mathlarger{\mathlarger{‎‎\sum}}_{\lambda_A,\lambda_A^\prime,\lambda_B,\lambda_B^\prime}^{}P_{AB(1,1)}\l(\lambda_A,\lambda_A^\prime,\lambda_B,\lambda_B^\prime\r)
		\times\Gamma_{A(1)}\l(\lambda_A,\lambda_A^\prime\r)
		\times\Gamma_{B(1)}\l(\lambda_B,\lambda_B^\prime\r) , \nonumber\\
		&=&\dfrac{1}{16\pi^2}\Big[1+ \alpha_A p_{i}^A c_{i}^a + \alpha_B p_{i}^B c_{i}^b + \alpha_A\alpha_B pp_{ij}^{AB}c_{i}^a c_{j}^b\Big],
	\end{eqnarray}
	where   $c_x$, $c_y$ and $c_z$ are angular functions of the daughters, i.e.,
	\begin{eqnarray}\label{eq:cor-defination}
		c_{x}^a = \sin\theta_a\cos\phi_a,~c_{y}^a = \sin\theta_a\sin\phi_a,~c_{z}^a = \cos\theta_a .
	\end{eqnarray}
	Here, the superscript $a$ on $c_i$ denote the daughter particle of the particle $A$.
One recovers Eq.~(\ref{eq:norm_dist-half}), the angular distribution of a spin-$1/2$ particle production and decay,
		by  integrating the angles of  $a(b)$ in Eq.~(\ref{eq:norm_dist-half-half}), i.e.,
		\begin{equation}
			\int_{\text{full}} d\Omega_{a(b)} \l( \dfrac{1}{\sigma}\dfrac{d^2\sigma}{d\Omega_{a}d\Omega_{b}} \r) = \dfrac{1}{\sigma}\dfrac{d\sigma}{d\Omega_{b(a)}}.
		\end{equation} 
	
	We know that individual polarizations $p_i^{A}$ and $p_i^{B}$  can be obtained from some asymmetries made by partially integrating the angular distributions~\cite{Rahaman:2016pqj,Rahaman:2020jll}. We write these asymmetries for $A$ in the following way, 
	\begin{eqnarray}\label{eq:Ap-int-rule}
		{\cal A}[p_{1}^{A}] &\equiv& \left( \int _{\theta_{a}=0}^{\pi }\int _{\phi_{a}=-\frac{\pi }{2}}^{\frac{\pi }{2}}
		-\int _{\theta_{a}=0}^{\pi }\int _{\phi_{a}=\frac{\pi }{2}}^{\frac{3 \pi }{2}}\right)
		d\Omega_{a}  \l(\frac{1}{\sigma}\frac{d\sigma}{d\Omega_{a}} \r) \nonumber \\
		&\equiv& \int_{a}^{p_1} d\Omega_{a}  \l(\frac{1}{\sigma}\frac{d\sigma}{d\Omega_{a}} \r)
		= \frac{1}{2}\alpha_{A} p_{1}^{A},\nonumber \\
		{\cal A}[p_{2}^{A}] &\equiv& \left(\int _{\theta_{a}=0}^{\pi }\int _{\phi_{a}=0}^{\pi }
		-\int _{\theta_{a}=0}^{\pi }\int _{\phi_{a}=\pi }^{2 \pi }\right)
		d\Omega_{a}  \l(\frac{1}{\sigma}\frac{d\sigma}{d\Omega_{a}} \r) \nonumber \\
		&\equiv& \int_{a}^{p_2} d\Omega_{a}  \l(\frac{1}{\sigma}\frac{d\sigma}{d\Omega_{a}} \r)
		= \frac{1}{2}\alpha_{A} p_{2}^{A},\nonumber \\
		{\cal A}[p_{3}^{A}] &\equiv&  \left( \int _{\theta_{a}=0}^{\frac{\pi }{2}}\int _{\phi_{a}=0}^{2 \pi }
		-\int _{\theta_{a}=\frac{\pi }{2}}^{\pi }\int _{\phi_{a}=0}^{2 \pi }\right)
		d\Omega_{a}  \l(\frac{1}{\sigma}\frac{d\sigma}{d\Omega_{a}} \r) \nonumber \\
		&\equiv& \int_{a}^{p_3} d\Omega_{a}  \l(\frac{1}{\sigma}\frac{d\sigma}{d\Omega_{a}} \r)
		= \frac{1}{2}\alpha_{A} p_{3}^{A}
	\end{eqnarray}
	to be used later for the asymmetry of spin correlations.  The $\int d\Omega_{b}$ integration is assumed to be performed in the above equation.  The asymmetries can also be represented as,
	\begin{equation}\label{eq:asym-pol-half}
		{\cal A}[p_{i}^{A}] = \dfrac{\sigma\l(c_{i}^a>0\r)-\sigma\l(c_{i}^a<0\r)}{\sigma\l(c_{i}^a>0\r)+\sigma\l(c_{i}^a<0\r)}
		= \frac{1}{2}\alpha_{A} p_{i}^{A}
	\end{equation}
	for numerical purpose. For the particle $B$, the polarization asymmetries are obtained by $A\to B,~a\to b$ in Eq.~(\ref{eq:Ap-int-rule}) and Eq.~(\ref{eq:asym-pol-half}).
	The correlations $pp_{ij}^{AB}$ can be obtained from some asymmetries made of partial integration  constructed from the individual partial integration in Eq.~(\ref{eq:Ap-int-rule}) as, 
	\begin{eqnarray}\label{eq:App-rule}
		{\cal A}[pp_{ij}^{AB}] &\equiv& \int_{a}^{p_i} d\Omega_{a}  \int_{b}^{p_j} d\Omega_{b} \l(\frac{1}{\sigma}\dfrac{d^2\sigma}{d\Omega_{a}d\Omega_{b}}\r) =\frac{1}{4}\alpha_A\alpha_B pp_{ij}^{AB} .
	\end{eqnarray}
	For example, the asymmetry for the correlation $pp_{12}^{AB}$ will be given by,
	\begin{eqnarray}\label{eq:App-12-rule}
		{\cal A}[pp_{12}^{AB}] &\equiv& \int_{a}^{p_1} d\Omega_{a}  \int_{b}^{p_2} d\Omega_{b} \l(\frac{1}{\sigma}\dfrac{d^2\sigma}{d\Omega_{a}d\Omega_{b}}\r), \nonumber\\
		&=&  \int _{\theta_{a}=0}^{\pi }\int _{\phi_{a}=-\frac{\pi }{2}}^{\frac{\pi }{2}} d\Omega_{a}
		\l[  \l(\int _{\theta_{b}=0}^{\pi }\int _{\phi_{b}=-\frac{\pi }{2}}^{\frac{\pi }{2}} 
		-\int _{\theta_{b}=0}^{\pi }\int _{\phi_{b}=\frac{\pi }{2}}^{\frac{3 \pi }{2}} \r) d\Omega_{b} \l(\frac{1}{\sigma}\dfrac{d^2\sigma}{d\Omega_{a}d\Omega_{b}}\r)  \r] \nonumber \\
		&&-\int _{\theta_{a}=0}^{\pi }\int _{\phi_{a}=\frac{\pi }{2}}^{\frac{3 \pi }{2}} d\Omega_{a}
		\l[  \l(\int _{\theta_{b}=0}^{\pi }\int _{\phi_{b}=-\frac{\pi }{2}}^{\frac{\pi }{2}} 
		-\int _{\theta_{b}=0}^{\pi }\int _{\phi_{b}=\frac{\pi }{2}}^{\frac{3 \pi }{2}} \r) d\Omega_{b} \l(\frac{1}{\sigma}\dfrac{d^2\sigma}{d\Omega_{a}d\Omega_{b}}\r)  \r]. \nonumber\\
	\end{eqnarray}
	For numerical purpose, the asymmetries for the spin correlations can also be represented as,
	\begin{eqnarray}\label{eq:asymmetries-half-half}
		{\cal A}[pp_{ij}^{AB}] &\equiv&   \dfrac{\sigma\l(c_{i}^a c_{j}^b>0\r)-\sigma\l(c_{i}^a c_{j}^b<0\r)}{\sigma\l(c_{i}^a c_{j}^b>0\r)+\sigma\l(c_{i}^a c_{j}^b<0\r)}
		=\frac{1}{4}\alpha_A\alpha_B pp_{ij}^{AB} .
	\end{eqnarray}
	In the next section, we discuss the formalism for spin-$1/2$ -- spin-$1$ correlations assuming $B$ to be a spin-$1$ particle.
	
	\subsubsection{Spin-$1/2$ -- spin-$1$ correlations}
	In the spin-$1/2$ -- spin-$1$ case, apart from the vector-vector correlations, we also have vector-tensor correlations, which we accommodate in
	the polarization-correlation density matrix with an outer product between a Pauli spin-$1/2$ matrix ($\tau$) and bilinear combination of the spin-$1$ matrices $S$. Thus the full spin correlated polarization density matrix of a system of spin-$1/2$ and spin-$1$ particles will be given by,
	\begin{eqnarray}\label{eq:spin-density-half-one}
		P_{AB\left(1,2\right)}\l(\lambda_A,\lambda_A^\prime,\lambda_B,\lambda_B^\prime\r) &=& \frac{1}{\l(2\times\frac{1}{2}+1\r)}\frac{1}{\l(2\times 1+ 1\r)}\Big[ \underbrace{\mathbb{I}_{2\times 2}\otimes \mathbb{I}_{3\times 3}}_{\mathbb{I}_{6\times 6}}  + \vec{p}^A\cdot\vec{\tau}\otimes \mathbb{I}_{3\times 3} + \dfrac{3}{2}\mathbb{I}_{2\times 2} \otimes \vec{p}^B\cdot\vec{S} \nonumber\\
		&+&\sqrt{\dfrac{3}{2}}  \mathbb{I}_{2\times 2} \otimes T_{ij}^B\big(S_iS_j+S_jS_i\big)
		+ pp_{ij}^{AB}\tau_i\otimes S_j + pT_{ijk}^{AB}\tau_i\otimes (S_jS_k+S_kS_j) \Big] ,\nonumber\\
		&&\hspace{0.45\textwidth}\l(i,j,k=x,y,z\r).
	\end{eqnarray} 
	Here, $p_i^{A/B}$ are the vector polarizations of $A/B$, $T_{ij}^B$ are the tensor polarizations of $B$ and  $pp_{ij}^{AB}$ are the vector-vector correlations as appeared in the previous section in Eq.~(\ref{eq:spin-density-half-half}). The $pT_{ijk}^{AB}$, representing vector-tensor correlations of $A$ and $B$, are the components of a third-rank tensor $pT^{AB}$.
	The tensor $pT^{AB}$ is symmetric and traceless under the last two indices, similar to the case of $T^B$, i.e., 
	\begin{eqnarray}\label{eq:constraints-half-one}
		&T_{(ij)}^B \equiv T_{ij}^B=T_{ji}^B, ~T_{ii}^B=0; \nonumber\\ 	
		&pT_{i(jk)}^{AB} \equiv pT_{ijk}^{AB}=pT_{ikj}^{AB}, ~ pT_{ijj}^{AB}=0 .
	\end{eqnarray}
	The parentheses $\l(\r)$ enclosing the indices $i,j$ represent that the tensor is symmetric under these two indices. 
	Here, we have a total of $3\times 5= 15$ independent vector-tensor correlations apart from $9$ vector-vector correlations, see Table~\ref{tab:num-param-half-one}.  
	\begin{table}[h!]
		\centering
		\begin{tabular}{|c|ccccc|c|}\hline
			Parameter & $p_{i}^A$ & $p_{i}^B$ & $T_{(ij)}^B$ & $pp_{ij}^{AB}$ & $pT_{i(jk)}^{AB}$ & Total \\ \hline
			Number of parameter & $3$ & $3$ &$5$ &$9$ &$15$ & $=35$\\ \hline
		\end{tabular}
		\caption{\label{tab:num-param-half-one} Number of independent polarization and spin correlation parameters in a spin-$1/2$ -- spin-$1$ pair of particles production.}
	\end{table}

	After combining the polarization density matrix in Eq.~(\ref{eq:spin-density-half-one}) with the normalized decay density matrices in Eqs.~(\ref{eq:decay-half}) for $A$ and (\ref{eq:decay-one}) for $B$ in accordance with Eq.~(\ref{eq:norm_dist}), we obtain the joint angular distribution of the decay products as,
	\begin{eqnarray}\label{eq:norm_dist-half-one}
		\dfrac{1}{\sigma}\dfrac{d^2\sigma}{d\Omega_{a}d\Omega_{b}}&=&\frac{2}{4\pi}\frac{3}{4\pi}
		\mathlarger{\mathlarger{‎‎\sum}}_{\lambda_A,\lambda_A^\prime,\lambda_B,\lambda_B^\prime}^{}P_{AB\left(1,2\right)}\l(\lambda_A,\lambda_A^\prime,\lambda_B,\lambda_B^\prime\r)
		\times\Gamma_{A\l(1\r)}\l(\lambda_A,\lambda_A^\prime\r)
		\times\Gamma_{B\l(2\r)}\l(\lambda_B,\lambda_B^\prime\r) , \nonumber\\
		&=&\dfrac{1}{16\pi^2}\Biggr[1+ \alpha_A p_{i}^A c_{i}^a + \frac{3}{2}\alpha_B p_{i}^B c_{i}^b + \sqrt{\frac{3}{2}} (1-3\delta_B)T_{ij}^B c_{i}^b c_{j}^b ~(i\ne j)\nonumber\\
		&+&\frac{1}{2}\sqrt{\frac{3}{2}}(1-3\delta_B)\underbrace{\left(T_{11}^B-T_{22}^B\right)}_{T_{11-22}^{B}}\underbrace{\left((c_{1}^b)^2-(c_{2}^b)^2\right)}_{\sin^2\theta_b\cos (2\phi_b)} \nonumber\\
		&+& \frac{1}{2}\sqrt{\frac{3}{2}}(1-3\delta_B) T_{33}^B\left(3(c_{3}^b)^2-1 \right)\nonumber\\
		&+& \alpha_A\alpha_B pp_{ij}^{AB} c_{i}^a c_{j}^b\nonumber\\
		&+&\alpha_A(1-3\delta_B)pT_{ijk}^{AB} c_{i}^a c_{j}^b c_{k}^b~(j\ne k)\nonumber\\
		&+&\frac{1}{2}\alpha_A(1-3\delta_B)\underbrace{\left(pT_{i11}^{AB}-pT_{i22}^{AB}\right)}_{pT_{i(11-22)}^{AB}} c_{i}^a\left((c_{1}^b)^2-(c_{2}^b)^2\right) \nonumber\\
		&+&\frac{1}{2} \alpha_A(1-3\delta_B) pT_{i33}^{AB} c_{i}^a \l(3(c_{3}^b)^2-1\r)
		\Biggr] .
	\end{eqnarray}
Here, systematization of indices on $T$ and $pT$ are not included.
	The independent vector-tensor correlations ($pT^{AB}$) are nine ($3\times 3$) $pT_{i(jk)}^{AB}~ (j\ne k)$, three $pT_{i(11-22)}^{AB}$, and three $pT_{i33}$. 
We recover angular distributions  in Eq.~(\ref{eq:norm_dist-half}) for spin-$1/2$ and Eq.~(\ref{eq:norm_dist-one}) for spin-$1$ particle production and decay after integrating the angles of $b$ and $a$ in Eq.~(\ref{eq:norm_dist-half-one}), respectively.
	In this case, the asymmetries for vector polarizations and vector-vector correlations are given by,
	\begin{eqnarray}\label{eq:asym-App-half-one}
		{\cal A}[p_i^A]=\frac{1}{2} \alpha_A p_i^A,~~
		{\cal A}[p_i^B]=\frac{3}{4} \alpha_B p_i^B,~~		
		{\cal A}[pp_{ij}^{AB}] =\frac{1}{4}\alpha_A\alpha_B pp_{ij}^{AB}.
	\end{eqnarray}
	The asymmetries for tensor polarizations ($T$) of $B$   are  given in appendix~\ref{app:tensor-integrand-rule}  in Eqs.~(\ref{eq:asym-T-12})-(\ref{eq:numericA-tensor-one}).
	The independent vector-tensor correlations $pT_{i(jk)}^{AB}$ can be obtained from the following asymmetries,
	\begin{eqnarray}\label{eq:asym-ApT-half-one}
		{\cal A}[pT_{i(jk)}^{AB}] &\equiv& \int_{a}^{p_i} d\Omega_{a}  \int_{b}^{T_{jk}} d\Omega_{b}
		\l(	\frac{1}{\sigma}\dfrac{d^2\sigma}{d\Omega_{a}d\Omega_{b}} \r) ~(j\ne k)\nonumber\\
		&=& 
		\dfrac{\sigma\l(c_i^a c_j^b c_k^b>0\r)-\sigma\l(c_i^a c_j^b c_k^b<0\r)}
		{\sigma\l(c_i^a c_j^b c_k^b>0\r)+\sigma\l(c_i^a c_j^b c_k^b<0\r)}\nonumber\\
		&=& \frac{2}{3\pi}\alpha_A(1-3\delta_B)pT_{i(jk)}^{AB},\\ 
		{\cal A}[pT_{i(11-22)}^{AB}] &\equiv& \int_{a}^{p_i} d\Omega_{a}  \int_{b}^{T_{11-22}} d\Omega_{b}
		\l(	\frac{1}{\sigma}\dfrac{d^2\sigma}{d\Omega_{a}d\Omega_{b}} \r) \nonumber\\
		&=& \dfrac{\sigma\l(c_i^a \l((c_{1}^b)^2-(c_{2}^b)^2\r)>0\r)-\sigma\l(c_i^a \l((c_{1}^b)^2-(c_{2}^b)^2\r)<0\r)}{\sigma\l(c_i^a \l((c_{1}^b)^2-(c_{2}^b)^2\r)>0\r)+\sigma\l(c_i^a \l((c_{1}^b)^2-(c_{2}^b)^2\r)<0\r)}  \nonumber \\
		&=& \frac{1}{3\pi}\alpha_A(1-3\delta_B) pT_{i(11-22)}^{AB},\\ 
		{\cal A}[pT_{i33}^{AB}] &\equiv& \int_{a}^{p_i} d\Omega_{a}  \int_{b}^{T_{33}} d\Omega_{b}
		\l(	\frac{1}{\sigma}\dfrac{d^2\sigma}{d\Omega_{a}d\Omega_{b}} \r) \nonumber\\
		&=&\dfrac{\sigma\l(c_i^a \sin(3\theta_b)>0\r)-\sigma\l(c_i^a \sin(3\theta_b)<0\r)}{\sigma\l(c_i^a \sin(3\theta_b)>0\r)+\sigma\l(c_i^a \sin(3\theta_b)<0\r)}\nonumber\\
		&=& \frac{3}{16}\alpha_A(1-3\delta_B)pT_{i33}^{AB}.
	\end{eqnarray}
	Here, the partial integration rules $\int_{b}^{T}$s are for the asymmetries related to tensor polarizations and they are given in appendix~\ref{app:tensor-integrand-rule}  in Eqs.~(\ref{eq:asym-T-12})-(\ref{eq:asym-T-33}). 
	In the next section, we discuss the formalism for spin-$1$ -- spin-$1$ correlations considering  $A$ also to be spin-$1$.

	\subsubsection{Spin-$1$ -- spin-$1$ correlations}
	For spin-$1$ -- spin-$1$ correlations, the polarization-correlation density matrix contains tensor-tensor correlations apart from the vector-vector and vector-tensor correlations. We accommodate the tensor-tensor correlations with an outer product between two bilinear combinations of spin-$1$ matrices ($S$). Thus the full spin correlated polarization density matrix for a pair of spin-$1$ particles will be given by,
	\begin{eqnarray}\label{eq:spin-density-one-one}
		P_{AB(2,2)}\l(\lambda_A,\lambda_A^\prime,\lambda_B,\lambda_B^\prime\r) &=& \frac{1}{\l(2\times 1 + 1\r)^2}\Bigg[\underbrace{\mathbb{I}_{3\times 3}\otimes \mathbb{I}_{3\times 3}}_{\mathbb{I}_{9\times 9}}  + \dfrac{3}{2}\vec{p}^A\cdot\vec{S}\otimes \mathbb{I}_{3\times 3} + \dfrac{3}{2}\mathbb{I}_{3\times 3} \otimes \vec{p}^B\cdot\vec{S} \nonumber\\
		&+&\sqrt{\dfrac{3}{2}} T_{ij}^A \big(S_iS_j+S_jS_i\big)\otimes \mathbb{I}_{3\times 3} 
		+\sqrt{\dfrac{3}{2}}  \mathbb{I}_{3\times 3} \otimes T_{ij}^B\big(S_iS_j+S_jS_i\big) \nonumber\\
		&+& pp_{ij}^{AB}S_i\otimes S_j + pT_{ijk}^{AB}S_i\otimes (S_jS_k+S_kS_j)\nonumber\\
		&+& Tp_{ijk}^{AB} (S_iS_j+S_jS_i)\otimes S_k \nonumber\\
		&+& TT_{ijkl}^{AB}\big(S_iS_j+S_jS_i\big)\otimes \big(S_kS_l+S_lS_k\big)
		\Bigg], ~\l(i,j,k,l=x,y,z\r).
	\end{eqnarray} 
	Here $TT_{ijkl}^{AB}$, components of a fourth-rank tensor, are the tensor-tensor correlations.  The $Tp_{ijk}^{AB}$ are tensor-vector correlations, which are identical to $pT_{kij}^{BA}$.  The $TT_{ijkl}^{AB}$ are symmetric and traceless under the first two and last two indices separately, i.e., 
	\begin{eqnarray}\label{eq:constrain-TT-one-one}
		TT_{(ij)(kl)}^{AB} \equiv TT_{ijkl}^{AB} =TT_{ijlk}^{AB} =TT_{jikl}^{AB} = TT_{jilk}^{AB},~~ 	TT_{(ij)kk}^{AB}=0=TT_{ii(jk)}^{AB}, 
	\end{eqnarray}
	which leaves us $5\times 5 =25$ independent tensor-tensor ($TT$) correlations, see Table~\ref{tab:num-param-one-one}. 
	\begin{table}[h!]
		\centering
		\begin{tabular}{|c|ccccccccc|}\hline
			Parameter &	$p_i^A$ & $p_i^B$& $T_{(ij)}^A$ & $T_{(ij)}^B$ & $pp_{ij}^{AB}$ & $pT_{i(jk)}^{AB}$& $pT_{i(jk)}^{BA}$ & $TT_{(ij)(kl)}^{AB}$  & Total \\ \hline
			Number of parameter &	$3$ & $3$ &$5$&$5$  &$9$ &$15$ &$15$ &$25$ & $=80$\\ \hline
		\end{tabular}
		\caption{\label{tab:num-param-one-one} Number of independent polarization and spin correlation parameters in a spin-$1$ -- spin-$1$ pair of particles production.}
	\end{table}
	
	After combining the polarization density matrix in Eq.~(\ref{eq:spin-density-one-one}) with the spin-$1$ normalized decay density matrix in Eq.~(\ref{eq:decay-one}) in accordance with Eq.~(\ref{eq:norm_dist}), we obtain the joint angular distribution of the decay products as,
	\begin{eqnarray}\label{eq:norm_dist-one-one}
		\dfrac{1}{\sigma}\dfrac{d^2\sigma}{d\Omega_{a}d\Omega_{b}}&=&\frac{3}{4\pi}\frac{3}{4\pi}
		\mathlarger{\mathlarger{‎‎\sum}}_{\lambda_A,\lambda_A^\prime,\lambda_B,\lambda_B^\prime}^{}P_{AB(2,2)}\l(\lambda_A,\lambda_A^\prime,\lambda_B,\lambda_B^\prime\r)
		\times\Gamma_{A(2)}\l(\lambda_A,\lambda_A^\prime\r)
		\times\Gamma_{B(2)}\l(\lambda_B,\lambda_B^\prime\r) \nonumber\\
		&=&\dfrac{1}{16\pi^2}\Bigg[1+ \frac{3}{2} \alpha_{A/B} p_{i}^{A/B} c_{i}^{a/b} 
		+ \sqrt{\frac{3}{2}} (1-3\delta_{A/B})T_{ij}^{A/B} c_{i}^{a/b} c_{j}^{a/b} ~(i\ne j)\nonumber\\
		&+&\frac{1}{2}\sqrt{\frac{3}{2}}(1-3\delta_{A/B})\underbrace{\left(T_{11}^{A/B}-T_{22}^{A/B}\right)}_{T_{11-22}^{A/B}}\underbrace{\left((c_1^{a/b})^2-(c_2^{a/b})^2\right)}_{\sin^2\theta_{a/b}\cos (2\phi_{a/b})} \nonumber\\
		&+& \frac{1}{2}\sqrt{\frac{3}{2}}(1-3\delta_{A/B}) T_{33}^{A/B}\left(3(c_3^{a/b})^2-1\right) \nonumber\\
		&+& \alpha_A\alpha_B  pp_{ij}^{AB }c_{i}^a c_{j}^b\nonumber\\
		&+&\alpha_A(1-3\delta_B)pT_{ijk}^{AB } c_{i}^a c_{j}^b  c_{k}^b 
		+\alpha_B(1-3\delta_A)pT_{ijk}^{BA} c_{i}^b c_{j}^a c_{k}^a~(j\ne k)\nonumber\\
		&+&\frac{1}{2}\alpha_A(1-3\delta_B)\underbrace{\left(pT_{i11}^{AB }-pT_{i22}^{AB }\right)}_{pT_{i(11-22)}^{AB}} c_{i}^a \l((c_1^b)^2-(c_2^b)^2\r) \nonumber\\
		&+&\frac{1}{2}\alpha_B(1-3\delta_A)\underbrace{\left(pT_{i11}^{BA }-pT_{i22}^{BA }\right)}_{pT_{i(11-22)}^{BA}} c_{i}^b \Big((c_1^a)^2-(c_2^a)^2\Big) \nonumber\\
		&+&\frac{1}{2} \alpha_A(1-3\delta_B) pT_{i33}^{AB} c_{i}^a \l(3(c_{3}^b)^2-1\r)
		+\frac{1}{2} \alpha_B(1-3\delta_A) pT_{i33}^{BA} c_{i}^b \l(3(c_{3}^a)^2-1\r)\nonumber\\
		&+&(1-3\delta_A)(1-3\delta_B) TT_{ijkl}^{AB} c_i^a c_j^a c_k^b c_l^b  ~(i\ne j,~k\ne l)\nonumber\\
		&+& \frac{1}{2}(1-3\delta_A)(1-3\delta_B)\underbrace{\l(TT_{ij11}^{AB}-TT_{ij22}^{AB}\r)}_{TT_{ij(11-22)}^{AB}}c_i^a c_j^a \l((c_1^b)^2-(c_2^b)^2\r)~(i\ne j) \nonumber\\
		&+& \frac{1}{2}(1-3\delta_A)(1-3\delta_B)\underbrace{\l(TT_{11ij}^{AB}-TT_{22ij}^{AB}\r)}_{TT_{(11-22)ij}^{AB}} c_i^b c_j^b \Big((c_1^a)^2-(c_2^a)^2\Big) ~(i\ne j) \nonumber\\
		&+&\frac{1}{2} (1-3\delta_A)(1-3\delta_B)TT_{ij33}^{AB} c_i^a c_j^a \l(3(c_3^b)^2-1\r) ~(i\ne j) \nonumber\\
		&+&\frac{1}{2} (1-3\delta_A)(1-3\delta_B)TT_{33ij}^{AB} c_i^b c_j^b \l(3(c_3^a)^2-1\r) ~(i\ne j) \nonumber\\
		&+& \frac{1}{4}(1-3\delta_A)(1-3\delta_B)\underbrace{(TT_{1111}^{AB}-TT_{1122}^{AB}-TT_{2211}^{AB}+TT_{2222}^{AB})}_{TT_{(11-22)(11-22)}^{AB}}\nonumber\\
		&\times& \Big((c_1^a)^2-(c_2^a)^2\Big) \l((c_1^b)^2-(c_2^b)^2\r)\nonumber\\
		&+& \frac{1}{4}(1-3\delta_A)(1-3\delta_B)\underbrace{(TT_{1133}^{AB}-TT_{2233}^{AB})}_{TT_{(11-22)33}^{AB}}\Big((c_1^a)^2-(c_2^a)^2\Big) \l(3(c_3^b)^2-1\r)\nonumber\\
		&+& \frac{1}{4}(1-3\delta_A)(1-3\delta_B)\underbrace{(TT_{3311}^{AB}-TT_{3322}^{AB})}_{TT_{33(11-22)}^{AB}}\l((c_1^b)^2-(c_2^b)^2\r) \l(3(c_3^a)^2-1\r)\nonumber\\
		&+& \frac{1}{4}(1-3\delta_A)(1-3\delta_B)TT_{3333}^{AB}\l(3(c_3^a)^2-1\r)\l(3(c_3^b)^2-1\r)
		\Bigg] .
	\end{eqnarray}
	In the above distribution, systematization of indices on $T$,  $pT$, and $TT$ are not included.
	One recovers the angular distribution  in  Eq.~(\ref{eq:norm_dist-one}) for spin-$1$ particle production and decay by integrating the angles of $a$ or $b$ in Eq.~(\ref{eq:norm_dist-one-one}).
	The independent tensor-tensor correlations ($TT^{AB}$) are nine $TT_{(ij)(kl)}^{AB}~(i\ne j, k\ne l)$,  three $TT_{(ij)(11-22)}^{AB}~(i\ne j)$, three  $TT_{(11-22)(ij)}^{AB}~(i\ne j)$, three $TT_{(ij)33}^{AB}~(i\ne j)$, three $TT_{33(ij)}^{AB}~(i\ne j)$,  one $TT_{(11-22)(11-22)}^{AB}$, one $TT_{(11-22)33}^{AB}$, one $TT_{33(11-22)}^{AB}$, and one $TT_{3333}^{AB}$. In this case, asymmetries for  the vector polarizations, vector-vector and vector-tensor correlations are given by,
	\begin{eqnarray}\label{eq:asym-App-ApT-one-one}
		{\cal A}[p_i^{A/B}]&=&\frac{3}{4} \alpha_{A/B} p_i^{A/B},\nonumber \\
		{\cal A}[pp_{ij}^{AB}] &=&  \frac{1}{4}\alpha_A\alpha_B pp_{ij}^{AB},\nonumber \\
		{\cal A}[pT_{i(jk)}^{AB}] 
		&=& \frac{2}{3\pi}\alpha_A(1-3\delta_B)pT_{i(jk)}^{AB} ~(j\ne k),\nonumber\\ 
		{\cal A}[pT_{i(jk)}^{BA}] 
		&=& \frac{2}{3\pi}\alpha_B(1-3\delta_A)pT_{i(jk)}^{BA} ~(j\ne k),\nonumber\\ 
		{\cal A}[pT_{i(11-22)}^{AB}]
		&=& \frac{1}{3\pi}\alpha_A(1-3\delta_B)pT_{i(11-22)}^{AB},\nonumber\\ 
		{\cal A}[pT_{i(11-22)}^{BA}]
		&=& \frac{1}{3\pi}\alpha_B(1-3\delta_A)pT_{i(11-22)}^{BA},\nonumber\\ 
		{\cal A}[pT_{i33}^{AB}]
		&=& \frac{3}{16}\alpha_A(1-3\delta_B)pT_{i33}^{AB},\nonumber\\
		{\cal A}[pT_{i33}^{BA}]
		&=& \frac{3}{16}\alpha_B(1-3\delta_A)pT_{i33}^{BA}.		
	\end{eqnarray}
	The asymmetries for the independent tensor-tensor correlations are given by, 
	\begin{eqnarray}\label{eq:Asymmetries-TT}
		{\cal A}[TT_{(ij)(kl)}^{AB}]&\equiv& \int_{a}^{T_{ij}}d\Omega_{a}  \int_{b}^{T_{kl}}d\Omega_{b} \l(\frac{1}{\sigma}\dfrac{d^2\sigma}{d\Omega_{a}d\Omega_{b}}\r) ~(i\ne j,~ k\ne l)\nonumber\\
		&=& \left(\frac{4}{3\pi}\right)^2 (1-3\delta_A)(1-3\delta_B) TT_{(ij)(kl)}^{AB},\\
		{\cal A}[TT_{(ij)(11-22)}^{AB}] &\equiv& \int_{a}^{T_{ij}}d\Omega_{a}  \int_{b}^{T_{11-22}}d\Omega_{b} \l(\frac{1}{\sigma}\dfrac{d^2\sigma}{d\Omega_{a}d\Omega_{b}}\r) ~ (i\ne j) \nonumber\\
		&=& \left(\frac{8}{9\pi^2}\right) (1-3\delta_A)(1-3\delta_B) TT_{(ij)(11-22)}^{AB},\\
		{\cal A}[TT_{(11-22)(ij)}^{AB}] 
		&\equiv&  \int_{a}^{T_{11-22}}d\Omega_{a}  \int_{b}^{T_{ij}}d\Omega_{b} \l(\frac{1}{\sigma}\dfrac{d^2\sigma}{d\Omega_{a}d\Omega_{b}}\r)  ~(i\ne j) \nonumber\\
		&=& \left(\frac{8}{9\pi^2}\right) (1-3\delta_A)(1-3\delta_B) TT_{(11-22)(ij)}^{AB} ,\\
		{\cal A}[TT_{(ij)33}^{AB}]&\equiv& \int_{a}^{T_{ij}}d\Omega_{a}  \int_{b}^{T_{33}}d\Omega_{b} \l(\frac{1}{\sigma}\dfrac{d^2\sigma}{d\Omega_{a}d\Omega_{b}}\r) ~ (i\ne j)  \nonumber\\
		&=&  \left(\frac{1}{2\pi}\right)^2 (1-3\delta_A)(1-3\delta_B) TT_{(ij)33}^{AB},\\
		{\cal A}[TT_{33(ij)}^{AB}]&\equiv&   \int_{a}^{T_{33}}d\Omega_{a}  \int_{b}^{T_{ij}}d\Omega_{b} \l(\frac{1}{\sigma}\dfrac{d^2\sigma}{d\Omega_{a}d\Omega_{b}}\r) ~ (i\ne j)  \nonumber\\
		&=& \left(\frac{1}{2\pi}\right)^2 (1-3\delta_A)(1-3\delta_B) TT_{33(ij)}^{AB} ~ (i\ne j),\\
		{\cal A}[TT_{(11-22)(11-22)}^{AB}] &\equiv& \int_{a}^{T_{11-22}}d\Omega_{a}  \int_{b}^{T_{11-22}}d\Omega_{b} \l(\frac{1}{\sigma}\dfrac{d^2\sigma}{d\Omega_{a}d\Omega_{b}}\r)  \nonumber\\
		&=& \left(\frac{4}{9\pi^2}\right) (1-3\delta_A)(1-3\delta_B) 
		TT_{(11-22)(11-22)}^{AB},\\ 
		{\cal A}[TT_{(11-22)33}^{AB}]&\equiv&  \int_{a}^{T_{11-22}}d\Omega_{a}  \int_{b}^{T_{33}}d\Omega_{b} \l(\frac{1}{\sigma}\dfrac{d^2\sigma}{d\Omega_{a}d\Omega_{b}}\r)   \nonumber\\
		&=& \l(\frac{1}{4\pi}\r) (1-3\delta_A)(1-3\delta_B) TT_{(11-22)33}^{AB},\\
		{\cal A}[TT_{33(11-22)}^{AB}] &\equiv&  \int_{a}^{T_{33}}d\Omega_{a}  \int_{b}^{T_{11-22}}d\Omega_{b} \l(\frac{1}{\sigma}\dfrac{d^2\sigma}{d\Omega_{a}d\Omega_{b}}\r)   \nonumber\\
		&=& \l(\frac{1}{4\pi}\r) (1-3\delta_A)(1-3\delta_B)  TT_{33(11-22)}^{AB},\\
		{\cal A}[TT_{3333}^{AB}] &\equiv& \int_{a}^{T_{33}}d\Omega_{a}  \int_{b}^{T_{33}}d\Omega_{b} \l(\frac{1}{\sigma}\dfrac{d^2\sigma}{d\Omega_{a}d\Omega_{b}}\r)  \nonumber\\
		&=& \l(\frac{9}{64}\r) (1-3\delta_A)(1-3\delta_B)  TT_{3333}^{AB}  .
	\end{eqnarray}
	For the numerical purpose, all  the independent ($25$) tensor-tensor correlation asymmetries can be obtained as, 
	\begin{equation}\label{eq:numericA-one-one}
		{\cal A}_{mn}[TT^{AB}]=\dfrac{\sigma\l({\cal C}_{m}^a {\cal C}_{n}^b>0\r)-\sigma\l({\cal C}_{m}^a {\cal C}_{n}^b<0\r)}{\sigma\l({\cal C}_{m}^a {\cal C}_{n}^b>0\r)+\sigma\l({\cal C}_{m}^a {\cal C}_{n}^b<0\r)},~~m,n\in[1,2,3,4,5], 
	\end{equation}
	where ${\cal C}_m$ are the combinations of $c_x,c_y,c_z$ given in Eq.~(\ref{eq:five-calC-app}) in appendix~\ref{app:tensor-integrand-rule}.

	The formalism we presented here for the spin correlations in all three cases, namely half-half, half-one and one-one spin cases, 
	gives the method to estimate all the spin correlations along with the polarizations from the normalized production density matrix as well as from the joint angular distribution by constructing several asymmetries. We test the formalism in some SM scattering processes in section~\ref{sec:sm-example} for all three scenarios for the purpose of sanity checking. 
	We will use some compact notations of the polarization and correlation parameters and their asymmetries for aesthetic visibility. These notations are $T_{x^2-y^2}=T_{11-22}$, $TT_{(mn)^2}=TT_{(mn)(mn)}$. For example, we will use $TT_{(xy)^2}=TT_{(12)(12)}$, $TT_{(x^2-y^2)^2}=TT_{(11-22)(11-22)}$, $TT_{(zz)^2}=TT_{3333}$ etc. 

	\subsection{Spin correlations in laboratory frame}\label{sec:corrltn-in-labframe}
	The values of the polarization parameters $p_i$ and $T_{ij}$ depend on the choice of the reference frame, and thus, the values of spin correlations do so. 
	The above formalism of spin polarization and correlations is based on the helicity frame, equivalent to the center-of-mass (CM) frame. For an $e^-$-$e^+$ collider,  CM frame and laboratory (Lab) frame are the same, while for a hadron collider such as the LHC, they are different due to the involvement PDFs. In a hadron collider, the polarization density matrix of a single particle receives an effective total rotation comprising boost and angular rotations, leaving the trace invariant going from CM to Lab frame. As a result, the polarization parameters $p_i$ and $T_{ij}$ get transformed as~\cite{Bourrely:1980mr,Arunprasath:2016tfq,Velusamy:2018ksp},
	\begin{eqnarray}\label{eq:cm-to-lab-pol}
		p_i^{\text{Lab}}&=& R_{ij}^Y(\omega)p_j^{\text{CM}},\nonumber\\
		T_{ij}^{\text{Lab}}&=& R_{ik}^Y(\omega)R_{jl}^Y(\omega)T_{kl}^{\text{CM}},
	\end{eqnarray} 
	where
	\begin{eqnarray}\label{eq:cm-to-lab-Tij}
		\cos \omega&=& \cos\theta_{\text{CM}} \cos\theta_{\text{Lab}} +\gamma_{\text{CM}} \sin\theta_{\text{CM}} \sin\theta_{\text{Lab}},\nonumber\\
		\sin \omega&=& \frac{m}{E}\left(\sin\theta_{\text{CM}} \cos\theta_{\text{Lab}} -\gamma_{\text{CM}} \cos\theta_{\text{\text{CM}}} \sin\theta_{\text{Lab}}\right).
	\end{eqnarray} 
	The spin correlation parameters $pp$, $pT$ and $TT$ of a system of two particles $A$ and $B$, thus, get transformed as,
	\begin{eqnarray}\label{eq:cm-to-lab-spin-corr}
		\l[pp_{ij}^{AB}\r]^{\text{Lab}} &=& R_{ik}^Y(\omega_A)R_{jl}^Y(\omega_B)\l[pp_{kl}^{AB}\r]^{\text{CM}},\nonumber\\
		\l[pT_{ijk}^{AB}\r]^{\text{Lab}} &=& R_{il}^Y(\omega_A)R_{jm}^Y(\omega_B) R_{kn}^Y(\omega_B) \l[pT_{lmn}^{AB}\r]^{\text{CM}},\nonumber\\
		\l[TT_{ijkl}^{AB}\r]^{\text{Lab}} &=& R_{im}^Y(\omega_A) R_{jn}^Y(\omega_A) R_{kp}^Y(\omega_B) R_{lq}^Y(\omega_B) \l[TT_{mnpq}^{AB}\r]^{\text{CM}}.
	\end{eqnarray} 
	Here, $R_{ij}^Y$ is the usual rotational matrix w.r.t. $y$-direction and 
	$\gamma_{\text{CM}}=1/\sqrt{1-\beta_{\text{CM}}^2}$ with $\beta_{\text{CM}}$ being boost of the CM 
	frame.
	The quantities $m$ and $E$ are the mass and energy of the particle in consideration, respectively.

	\subsection{Examples of spin correlations in the SM}\label{sec:sm-example}
	\begin{figure}[ht!]
		\centering
		\includegraphics[width=0.496\textwidth]{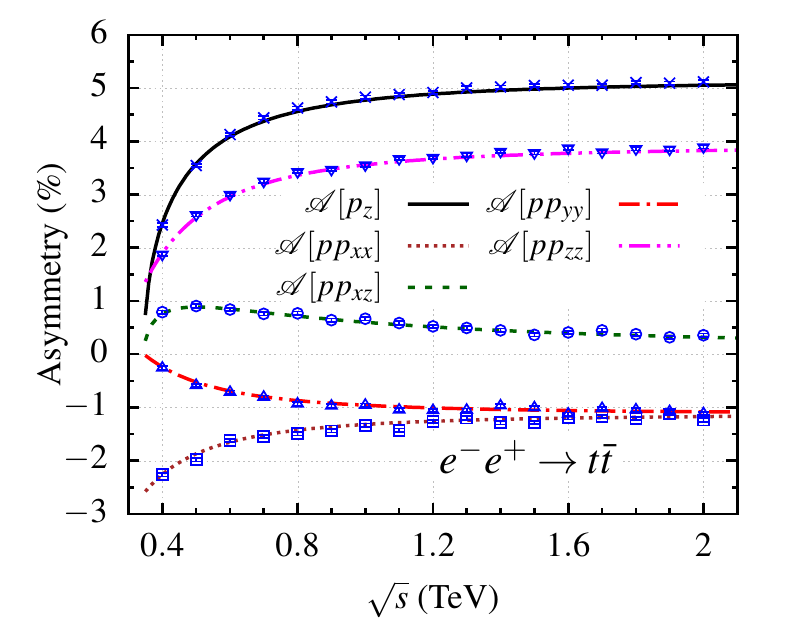}
		\includegraphics[width=0.496\textwidth]{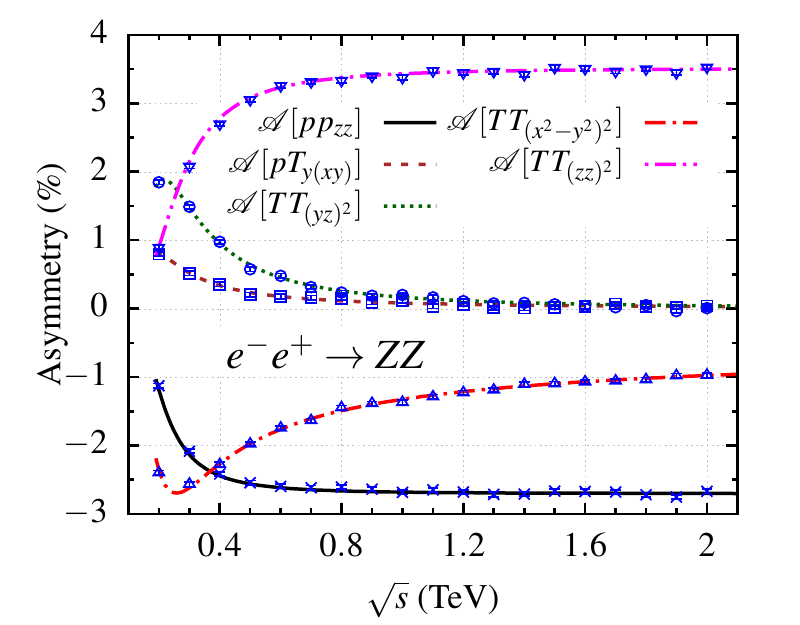}
		\includegraphics[width=0.496\textwidth]{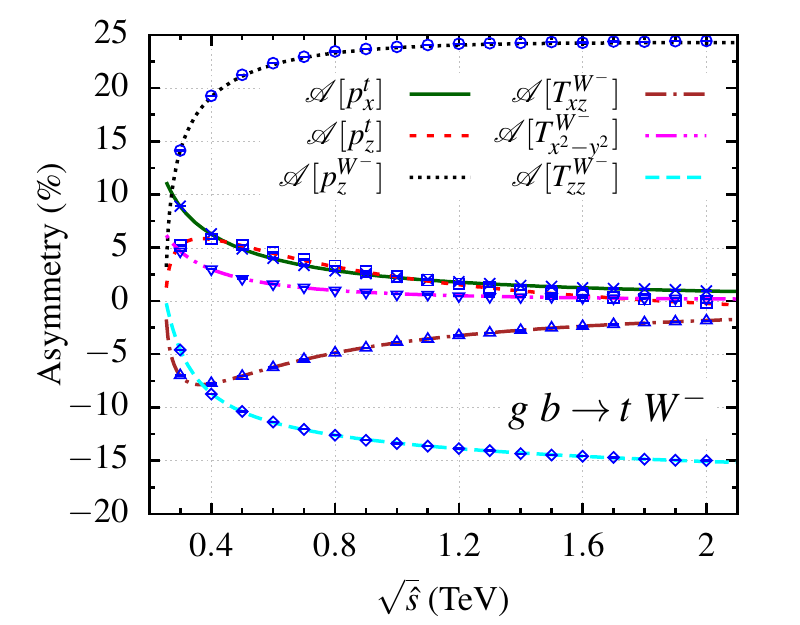}
		\includegraphics[width=0.496\textwidth]{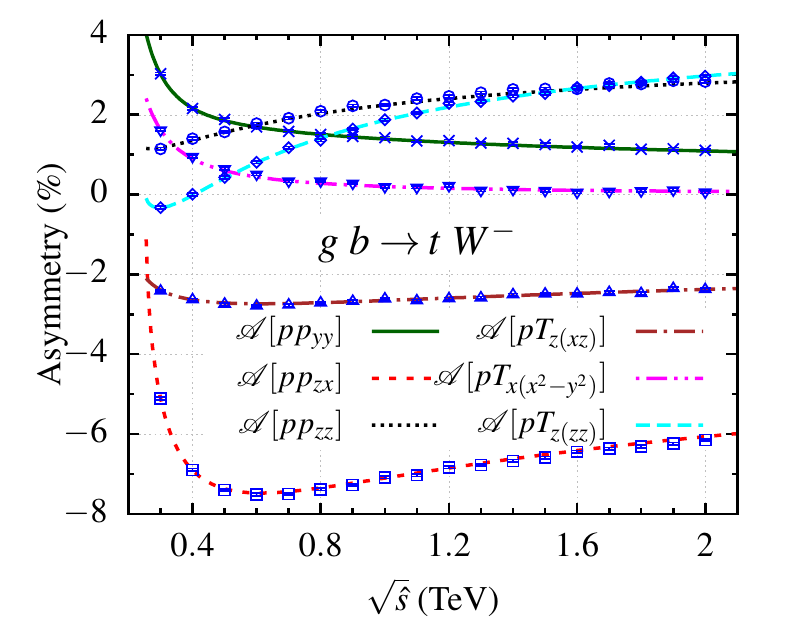}
		\caption{\label{fig:Asym-sanity-check}. The SM values of the asymmetries for spin correlations and polarizations are shown as a function of center-of-mass energy (c.m.e) for  $e^-e^+\to t\bar{t}$ in the {\em left-top-panel}, $e^-e^+\to ZZ$ in the  {\em right-top-panel} and  $gb \to tW^-$ in the {\em bottom-panel}. The data points with errorbars correspond to $10^7$ events generated in \MGvATNLO.} 
	\end{figure}
	As a demonstration of the formalism discussed above 
	for the spin correlations along with polarizations, we choose three
	processes namely  $e^-e^+\to t\bar{t}$, $gb \to tW^-$ and $e^-e^+\to ZZ$ as examples of all three scenarios, i.e.,  half-half ($t\bar{t}$),   half-one ($tW^-$) and one-one ($ZZ$)   spin systems.  The correlation and polarization parameters are constructed at the production level of the particles using the polarization-correlation matrices in Eqs.~(\ref{eq:spin-density-half-half}),  (\ref{eq:spin-density-half-one}), (\ref{eq:spin-density-one-one}) 
	 as well as in their decay level using the normalized angular distributions of the decay products (Eqs.~ (\ref{eq:norm_dist-half-half}),  (\ref{eq:norm_dist-half-one}),  (\ref{eq:norm_dist-one-one})) using the partial integration  assuming on-shell production and decay of the mother particles. 
	We calculate all the asymmetries related to spin correlations and polarizations of the particles at the production level using the helicity amplitude technique  as a function of  center-of-mass energy (c.m.e) in all  three processes. At first, the polarization and correlations parameters are  obtained with the method   explained in appendix~\ref{app:corr-from-production}. The asymmetries are then obtained with the appropriate coefficients given above. 

	We find all the $CP$-odd polarization and correlation parameters, i.e., all the parameters with only one $y$- suffix, e.g., $p_y$, $T_{xy}$, $pp_{yz}$, $pT_{x(yz)}$, $TT_{(xy)zz}$ to be zero as they should be for the SM being $CP$ conserving. However, correlations appearing with two  $y$- suffix, e.g., $pp_{yy}$, can be non zero in general. 
	
	In the $t\bar{t}$ production process, all the polarization and correlations except the $CP$-odd ones are non vanishing. Furthermore, both $t$ and $\bar{t}$ have same values of  $p_x$ and opposite values of $p_z$, i.e., $p_x^t=p_x^{\bar{t}}$ and $p_z^t=-p_z^{\bar{t}}$.
	The correlations $pp_{xz}$ and $pp_{zx}$ are equal and opposite because of the nature of individual $p_x$ and $p_z$.  
	
	In the $tW^-$ production process, all the $CP$-even polarization and correlations are non-vanishing. There are no relations among the polarizations and correlations here,  as the final state particles are entirely different.

	In the $ZZ$ production process, final state being symmetric, there are only eight independent polarizations and thirty six independent correlations\footnote{ Independent parameters are $3$ piece $p$, $5$ piece $T$, $6$ independent $pp$, $15$ independent $pT$ and $15$ independent $TT$.}.
	Owing to the initial state symmetry,  there are only three non-vanishing polarizations 
	($p_x$, $T_{x^2-y^2}$, $T_{zz}$), three non-vanishing vector-vector correlations (diagonals only, i.e., $pp_{ii},~i=x,y,z$), three non-vanishing vector-tensor 
	correlations ($pT_{y(xy)}$, $pT_{x(x^2-y^2)}$, $pT_{xzz}$), and six non-vanishing tensor-tensor correlations ($TT_{(xy)^2}$, $TT_{(xz)^2}$, $TT_{(yz)^2}$, $TT_{\l(x^2-y^2\r)^2}$, $TT_{(x^2-y^2)zz}$, $TT_{(zz)^2}$) totaling $3+12=15$ non-vanishing polarization and correlations. Note that the correlation $TT_{(xz)^2}$
	is non-vanishing even the  $T_{xz}$ vanishes. 
	Besides these, correlation of two $CP$ odd polarizations  ($pp_{yy}$, $pT_{y(xy)}$, $TT_{(xy)^2}$, $TT_{(yz)^2}$) came out to be non-vanishing.

	We further calculate all the asymmetries related to the polarizations and correlations using the angular distributions of the daughter particles in each production process by generating large number of events ($10^7$) in \MGvATNLO~({\tt mg5\_aMC}) Monte-Carlo event generator.  The full processes, including decay,  are chosen as follows,
	\begin{eqnarray}\label{eq:decay-process-sm}
		e^-e^+\to t\bar{t}&:&~~ t\to b W^+,~ W^+\to l^+\nu_l;~ \bar{t}\to \bar{b} W^-,~ W^-\to l^- \bar{\nu_l},\nonumber\\
		gb \to tW^-&:&~~ t\to b W^+,~ W^+\to l^+\nu_l;~ W^-\to l^- \bar{\nu_l} ,\nonumber\\
		e^-e^+\to ZZ&:&~~ Z\to e^-e^+,~ Z\to \mu^+\mu^- .
	\end{eqnarray}
	We use the $b$ (for $t$) and $\bar{b}$ (for $\bar{t}$)  angular distributions in $e^-e^+\to t\bar{t}$ process,  $b$ (for $t$)  and $l^-$ (for $W^-$) angular distributions in $gb \to tW^-$ process, and  $e^-$ (for $Z_1$) and $\mu^-$ (for $Z_2$) angular distributions in $e^-e^+\to ZZ$ process.  For the top ($t$) polarization, one can also use the angular distributions of secondary lepton ($l$) with different analyzing power~\cite{Godbole:2006tq}.
	The SM values for some chosen  asymmetries (in~$\%$) (away from zero) related to spin correlations and polarizations are shown
	in Fig.~\ref{fig:Asym-sanity-check} as a function of  c.m.e  ($\sqrt{s}$) in all three processes.
	The values for the same asymmetries from the Monte-Carlo simulation are shown with data points with an errorbar corresponding to $10^7$ events.
	One easily finds an excellent agreement between the  analytical values shown by lines  and the  Monte-Carlo simulated values shown by the points for a range of c.m.e. 
   The asymmetries tend to saturate at some value as the energy increases in all three cases. 
	The reasons are the following:  The polarization and spin-correlation parameters are ratios of two quantities (cross sections), both depending on the velocity, which approaches unity as the energy increases. As a result, the polarization and spin-correlation parameters saturate to some values. We note that different cuts such as transverse momentum ($p_T$), pseudorapidity ($\eta$), etc., which are necessary for realistic scenarios, on the daughter particles will reduce the full angular phase-space. In that case, one can not recover the actual polarization and spin-correlation parameters from the asymmetries discussed above. Nevertheless, the asymmetries will be related to the polarizations and spin-correlations in the process, and they can be used to study  BSM physics.
	We note that, although the spin density matrices assume on-shell production and decay, the finite width effect  included in the \MGvATNLO~(see Eq.~(\ref{eq:decay-process-sm})) simulation gives the identical results.
	In the next section, we investigate how these polarizations and correlations are affected by BSM physics and how the correlations perform over the polarizations   in probing BSM physics. 
	
	\section{New physics effect  on spin correlations and polarizations}\label{sec:bsm-examples}
	We investigate the effect of possible new physics on the spin correlations and polarizations   in  $e^-e^+\to t\bar{t}$,  $gb \to tW^-$ and $u\bar{d}\to ZW^+$   processes  as example of all three scenarios, namely half-half ($t\bar{t}$), half-one ($tW^-$)  and one-one ($ZW^+$)    spin systems.
	We consider $\gamma t\bar{t}$,   $gt\bar{t}$ and  $W^+W^-Z$  anomalous couplings as  examples of new physics for $t\bar{t}$,   $tW^-$ and $ZW^+$ production processes, respectively. The BSM Lagrangians for the $\gamma t\bar{t}$~\cite{AguilarSaavedra:2012vh}, $gt\bar{t}$~\cite{Aguilar-Saavedra:2018ggp} and  $W^+W^-Z$~\cite{Hagiwara:1986vm}   anomalous couplings are given by,
	\begin{eqnarray}\label{eq:Lag-BSM}
		{\cal L}_{\gamma t\bar{t}} &=&  -i e \bar{t} \frac{\sigma^{\mu\nu} q_\nu}{m_t}
		\left( d_V^\gamma + i d_A^\gamma \gamma_5 \right) t A_\mu , \nonumber \\
		{\cal L}_{gt\bar{t}} &=&    -\frac{g_s}{m_t} \bar{t} \sigma^{\mu\nu} 
		\left( d_V^g + i d_A^g \gamma_5 \right)\frac{\lambda^a}{2} t G_{\mu\nu}^a ,\nonumber\\
		{\cal L}_{WWZ} &=&-\frac{i e \cos\theta_W}{\sin\theta_W}\l[
		\frac{\lambda^Z}{m_W^2}W_\mu^{+\nu}W_\nu^{-\rho}Z_\rho^{\mu}
		+\frac{\wtil{\lambda^Z}}{m_W^2}W_\mu^{+\nu}W_\nu^{-\rho}\wtil{Z}_\rho^{\mu} \r] .
	\end{eqnarray}
	Here, $q$ is the four-momentum transfer of the photon in $\mathcal{L}_{\gamma t\bar{t}}$;    
	$G_{\mu\nu}^a $ are the gluon field strength tensor; $\lambda^a$ are the Gell-Mann matrices; $g_s$ is the strong coupling constant; $m_t$ is the top quark mass;  $\theta_W$ is the weak mixing angle: $W_{\mu\nu}^\pm = \partial_\mu W_\nu^\pm - \partial_\nu W_\mu^\pm$, 
	$Z_{\mu\nu} = \partial_\mu Z_\nu - \partial_\nu Z_\mu$, 
	$\wtil{Z}^{\mu\nu}=1/2\epsilon^{\mu\nu\rho\sigma}Z_{\rho\sigma}$. The $\gamma t\bar{t}$,   $gt\bar{t}$ and  $W^+W^-Z$ vertices can have more anomalous couplings, but we restrict to only those in Eq.~(\ref{eq:Lag-BSM}) for simplicity.  
	All the couplings $d_{V,A}^{\gamma/g}$, $\lambda^Z$ and $\wtil{\lambda^Z}$ are zero in the SM; $d_{V}^{\gamma/g}$ and $\lambda^Z$ are $CP$-even; $d_{A}^{\gamma/g}$ and $\wtil{\lambda^Z}$ are $CP$-odd. 
	The coupling $d_V^\gamma$ ($d_V^g$)  corresponds to top quark magnetic (chromomagnetic) dipole moment, where as  $d_A^\gamma$ ($d_A^g$) correspond to  electric (chromoelectric) dipole moment.

	\paragraph{Effect of anomalous  $\gamma t\bar{t}$ couplings   in {\boldmath $e^-e^+\to t\bar{t}$}   :} 
	\begin{figure}[ht!]
		\centering
		\includegraphics[width=0.65\textwidth]{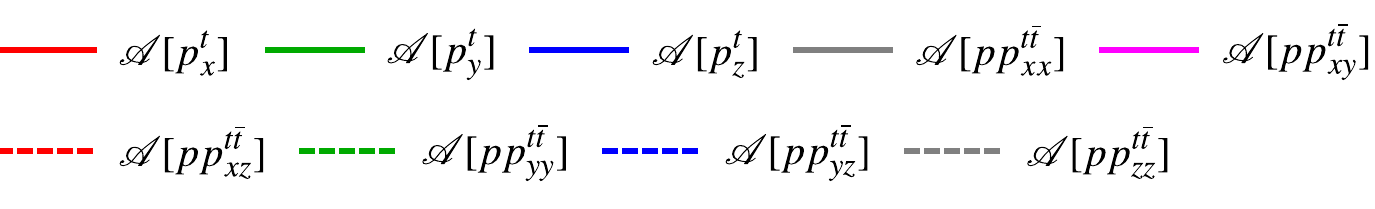}\\
		\includegraphics[width=0.49\textwidth]{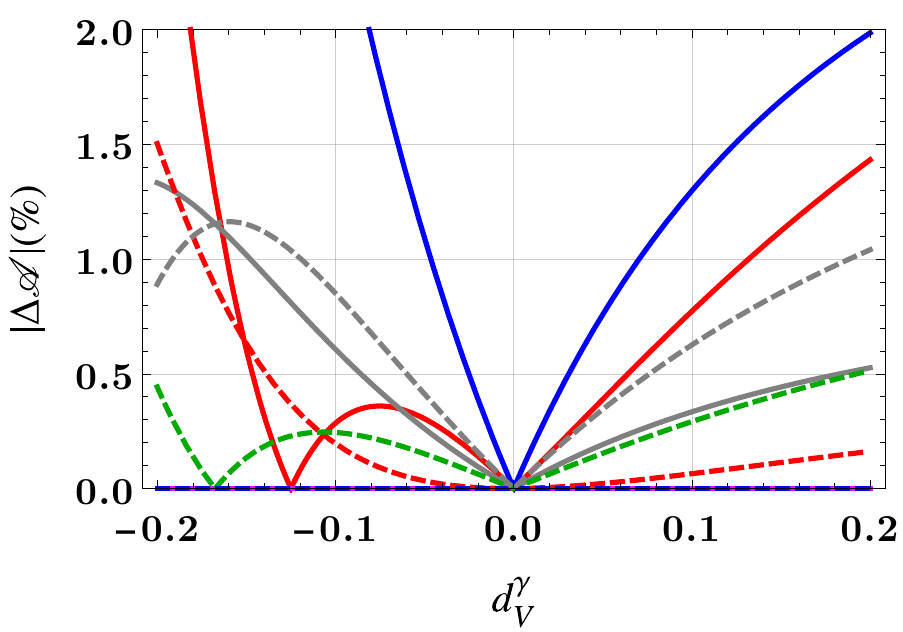}
		\includegraphics[width=0.49\textwidth]{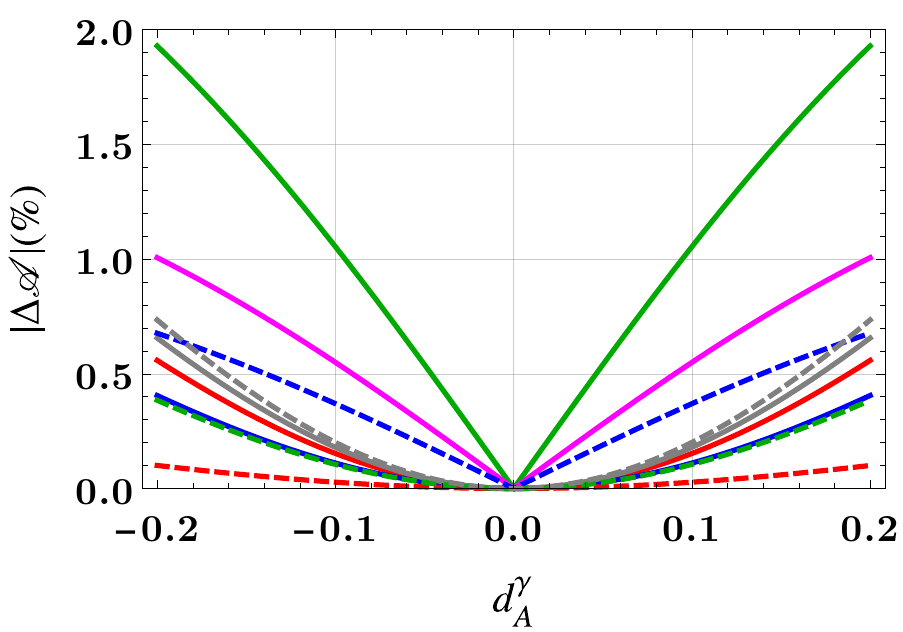}
		\includegraphics[width=0.49\textwidth]{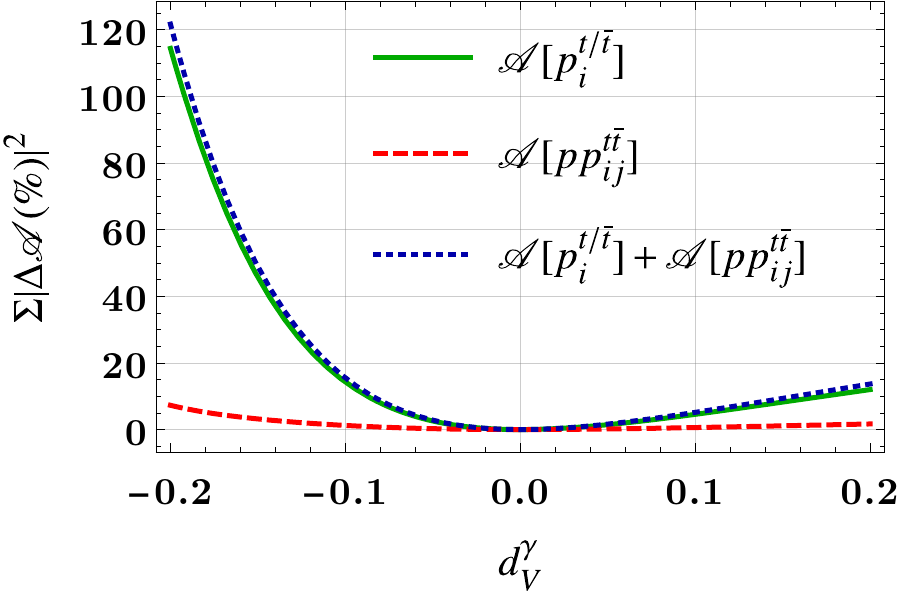}
		\includegraphics[width=0.49\textwidth]{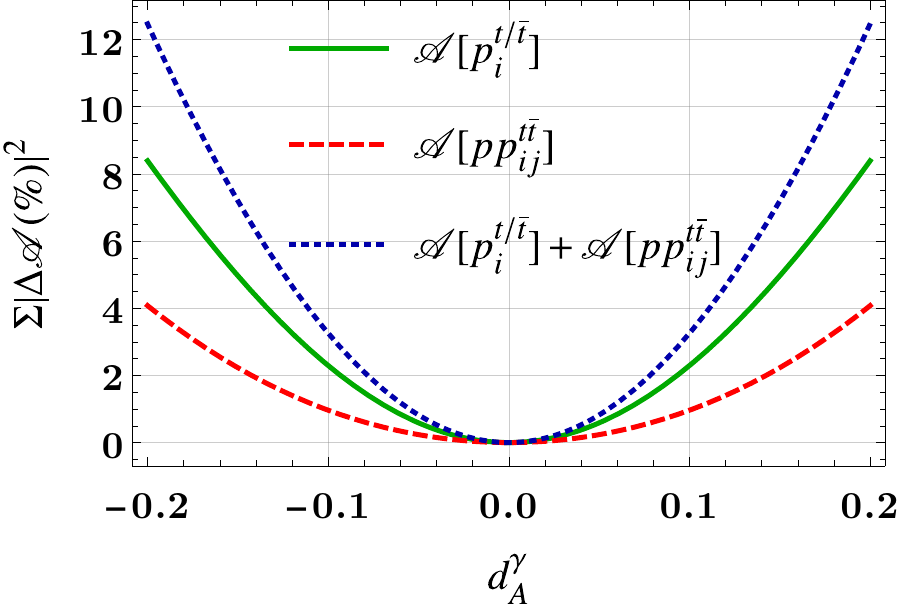}
		\caption{\label{fig:dA-tt} The absolute deviation of the asymmetry from the SM, i.e., $|\Delta {\cal A}|=|{\cal A}_{SM+BSM}-{\cal A}_{SM}|$ and their square sum, i.e., $\sum|\Delta {\cal A}|^2$ are shown as a function of anomalous couplings in the process  $e^-e^+\to t\bar{t}$  at $\sqrt{s}=500$ GeV.} 
	\end{figure}
	We calculate  all the polarizations of $t$ and $\bar{t}$ and their spin
	correlations in $e^-e^+\to t\bar{t}$, analytically from the production process, as a function of the anomalous couplings $d_V^\gamma$ and $d_A^\gamma$  for a fixed c.m.e of $\sqrt{s}=500$ GeV in the same decay channel as considered in the SM example given in Eq.~(\ref{eq:decay-process-sm}). 
	The $CP$-odd coupling $d_A^\gamma$ appears linearly on the numerator of  $CP$-odd parameters ($p_y^{t/\bar{t}}$, $pp_{ij},~(i/j=y)$) and only quadratically on the diagonal correlations $pp_{ii}$. The $CP$-even coupling $d_V^\gamma$
	appears linearly on the numerators of $CP$-even polarizations ($p_x^{t/\bar{t}}$, $p_z^{t/\bar{t}}$) and off diagonal correlations $pp_{ij},~(i\ne j)$, while the numerators of diagonal correlations ($pp_{ii}$) have linear as well as quadratic contribution of $d_V^\gamma$. The polarization asymmetries for $t$ and $\bar{t}$ are identical except ${\cal A}[p_x]$ being opposite.  
	We calculate the deviations of all the asymmetries from the SM, i.e., $\Delta {\cal A}={\cal A}_{SM+BSM}-{\cal A}_{SM}$  as  functions of the couplings and show  their absolute values ($|\Delta {\cal A}|$) in percentage ($\%$)   in 
		Fig.~\ref{fig:dA-tt} in the {\em top-row} to see their relative strengths. 
	Here, $|\Delta {\cal A}[p_i^t]|=|\Delta {\cal A}[p_i^{\bar{t}}]|$ and $|\Delta {\cal A}[pp_{ij}^{t\bar{t}}]|=|\Delta {\cal A}[pp_{ji}^{t\bar{t}}]|$. In Fig.~\ref{fig:dA-tt}  {\em top-row}, only one asymmetry is shown in the case of degenerate asymmetries.  
	The  $|\Delta {\cal A}|$ can be considered as sensitivity if we assume $1\%$
	errors for the asymmetries, which corresponds to $10^4$ events in the SM.
	We see that, polarization asymmetries  ${\cal A}[p_z^{t/\bar{t}}]$, ${\cal A}[p_x^{t/\bar{t}}]$ show large deviation for  small values of  $d_V^\gamma$ and 
	${\cal A}[p_y^{t/\bar{t}}]$ for small values of $d_A^\gamma$. The correlation asymmetries   also show similar behavior, especially for $d_A^\gamma$.  We compare the correlation asymmetries and the polarization asymmetries by taking quadratic sum of the 
	$|\Delta {\cal A}|$, i.e., $\sum_i |\Delta {\cal A}[p_i]|^2$ and  $\sum_{ij} |\Delta {\cal A}[pp_{ij}]|^2$.  These quadratic sums are equivalent to $\chi^2$
	for $10^4$ events in the SM. In Fig.~\ref{fig:dA-tt} {\em bottom-row},  we show the  $\sum|\Delta {\cal A}|^2$ for polarization asymmetries, correlation asymmetries and their sum. 
	The correlation asymmetries offer significant improvement over the polarization asymmetries in probing the anomalous couplings,   especially for the $CP$-odd coupling $d_A^\gamma$.

	\paragraph{Effect of anomalous $gt\bar{t}$ couplings in {\boldmath $gb\to tW^-$} : }
	\begin{figure}[ht!]
		\centering
		\includegraphics[width=1\textwidth]{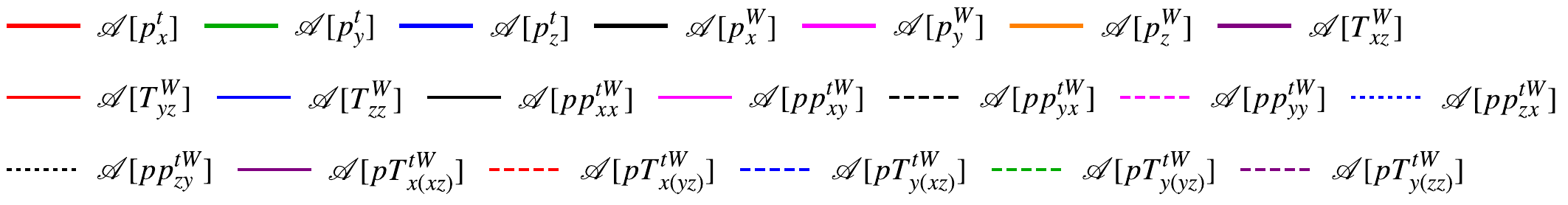}	
		\includegraphics[width=0.496\textwidth]{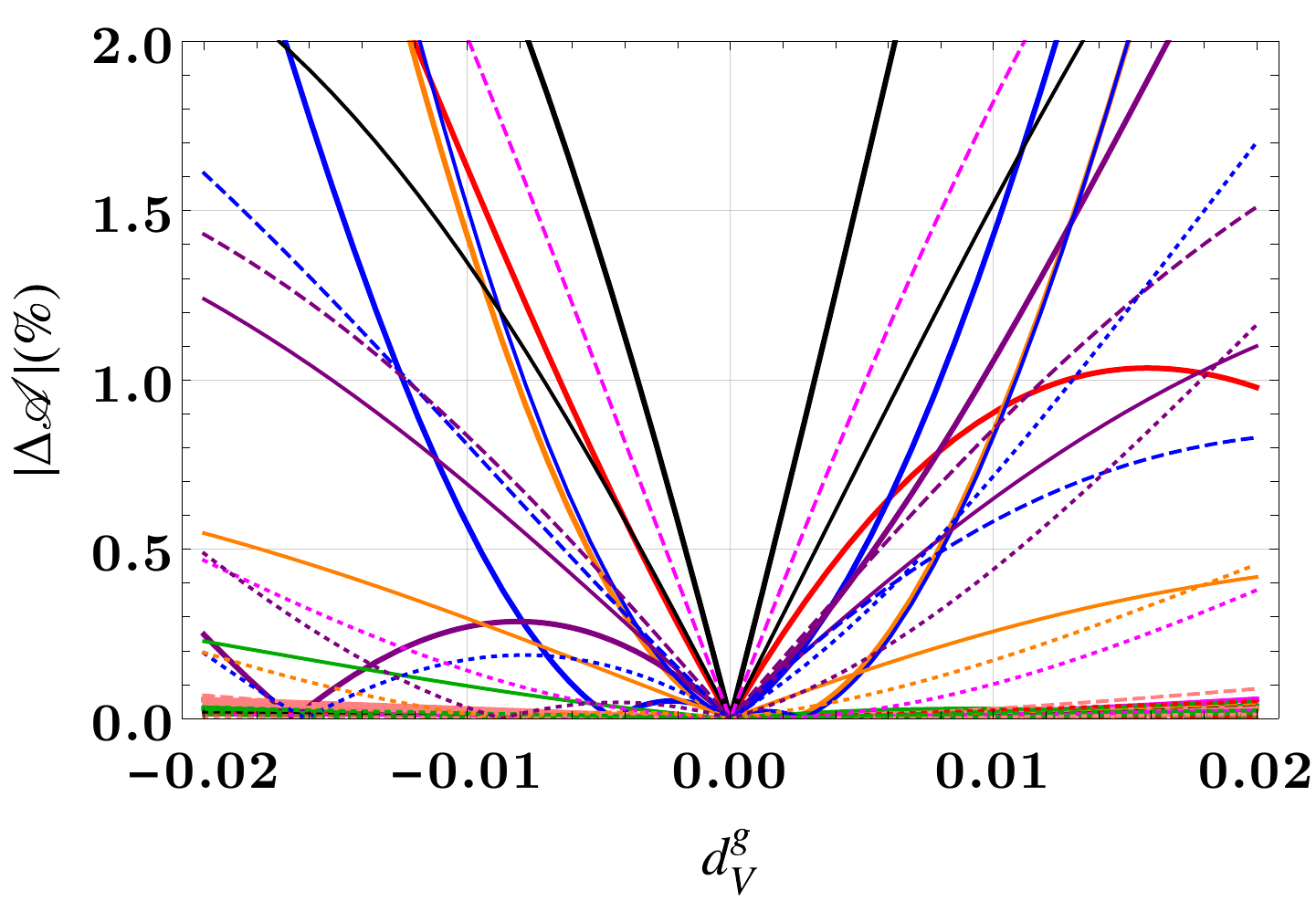}
		\includegraphics[width=0.496\textwidth]{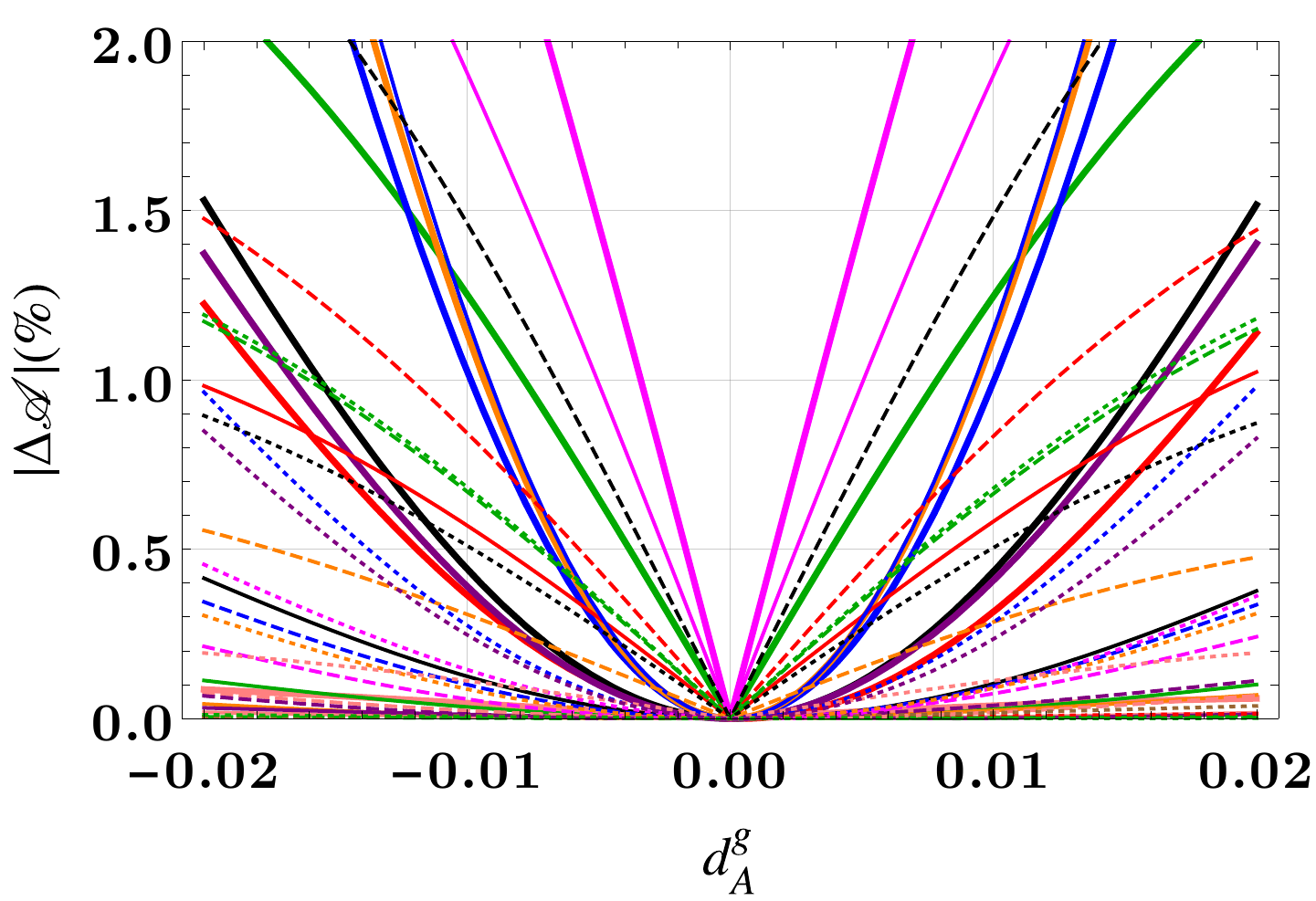}
		\includegraphics[width=0.496\textwidth]{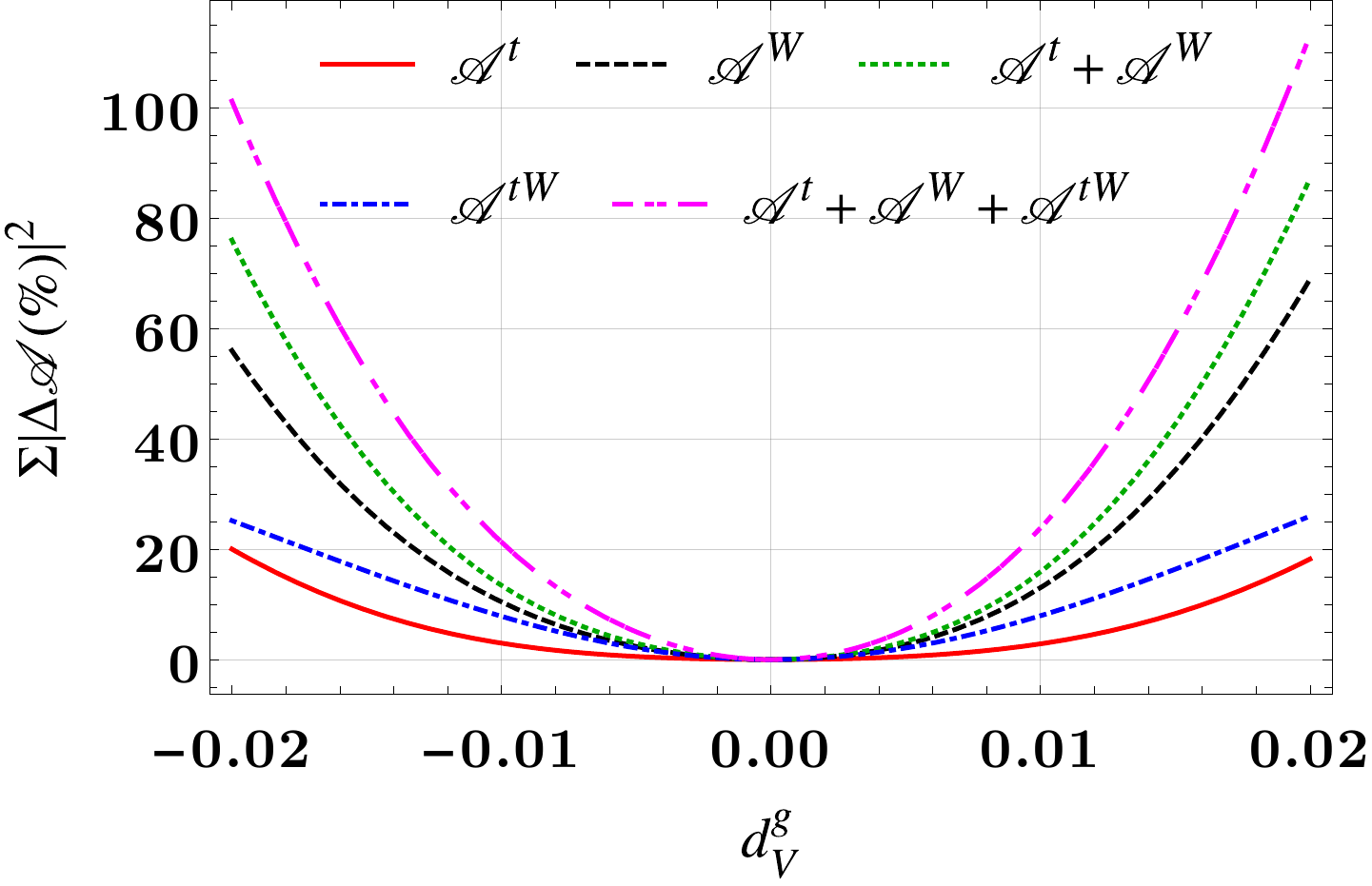}
		\includegraphics[width=0.496\textwidth]{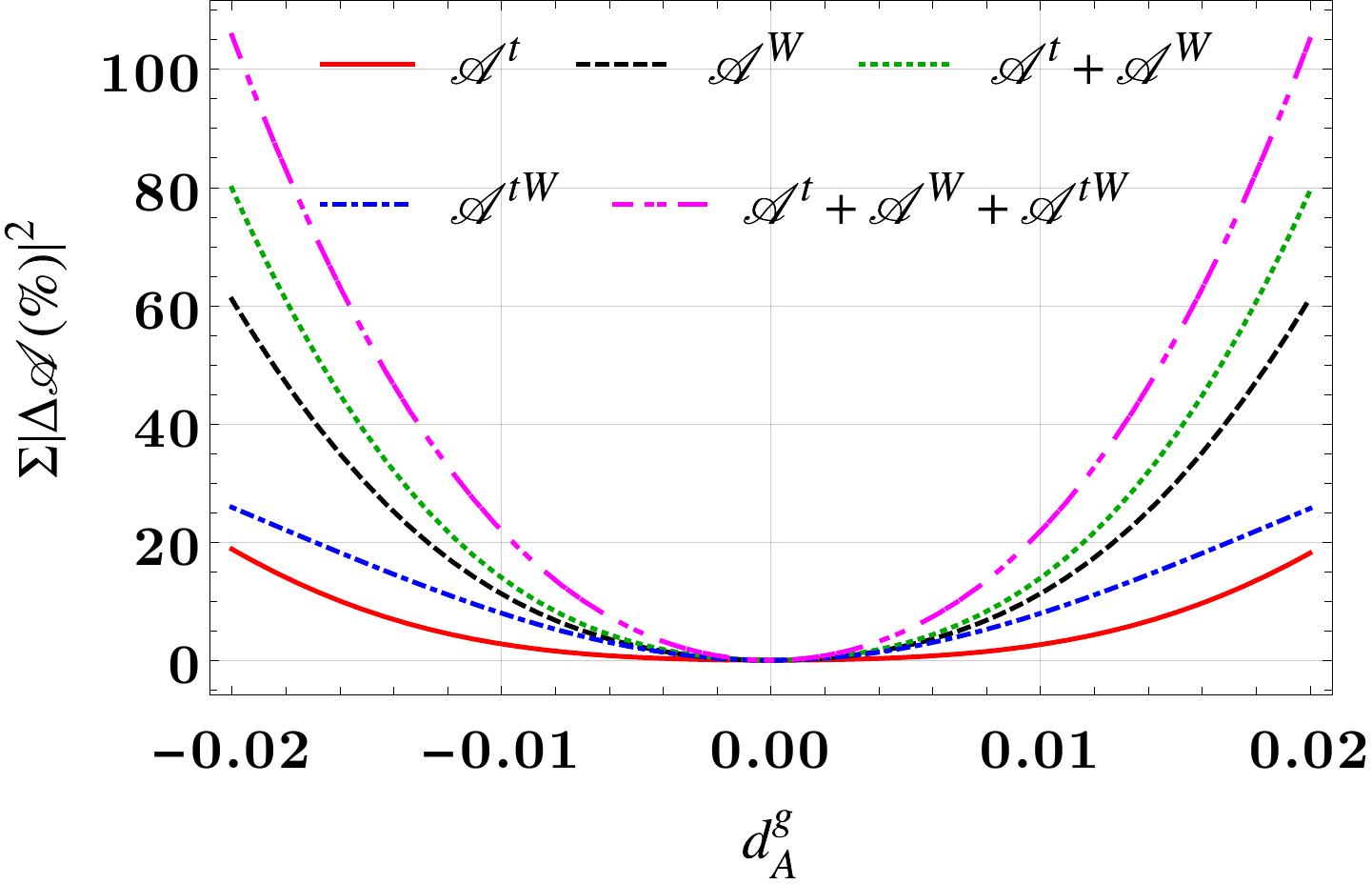}	
		\caption{\label{fig:dA-tW} The absolute deviation of the asymmetry from the SM  ($|\Delta {\cal A}|$) and their square sum  ($\sum|\Delta {\cal A}|^2$) are shown as a function of anomalous couplings in the process  $gb\to tW^-$  at $\sqrt{\hat{s}}=1$ TeV.} 
	\end{figure}
	In $gb\to tW^-$ partonic process, we choose the same decay channel chosen for the SM example in sec.~\ref{sec:sm-example} given in Eq.~(\ref{eq:decay-process-sm}). 
	In this case, we use decay distribution of the daughters of $t$ and $W^-$ to obtain their polarization and correlation asymmetries by generating large number of events ($10^6$) in {\tt mg5\_aMC} for a range of couplings $d_{V/A}^g$  at partonic $\sqrt{\hat{s}}=1$ TeV .

	In this process, all the polarization and correlation asymmetries are entirely different from each other.  
	In Fig.~\ref{fig:dA-tW} {\em top-row}, we show the absolute deviation of the asymmetries from the SM ($|\Delta {\cal A}|$) as a function of couplings for all the asymmetries for completeness. The legends are shown only for the asymmetries having $|\Delta {\cal A}| >1$ within the range of couplings shown. 
	There are many correlation asymmetries along with polarization asymmetries that  show large $|\Delta {\cal A}|$ for small values of couplings. 
	For example, along with  the polarization asymmetries ${\cal A}[p_x^W]$ and ${\cal A}[p_x^t]$, the correlation asymmetries ${\cal A}[pp_{yy}^{tW}]$ and ${\cal A}[pp_{xx}^{tW}]$ also show large $\Delta {\cal A}$ for small values of $d_V^g$, see Fig.~\ref{fig:dA-tW} {\em left-top-panel}. Similarly for the $CP$-odd coupling $d_A^g$, shown in Fig.~\ref{fig:dA-tW} {\em right-top-panel}, the correlation asymmetries ${\cal A}[pp_{xy}^{tW}]$ and ${\cal A}[pp_{yx}^{tW}]$ along with the polarization asymmetries ${\cal A}[p_y^W]$ and ${\cal A}[p_y^t]$     have large $\Delta {\cal A}$ for small values of coupling. 
	Thus it is expected that the correlation asymmetries will improve over the polarization asymmetries in $\sum|\Delta {\cal A}|^2$, which are shown in Fig.~\ref{fig:dA-tW}
		in the {\em bottom-row} for both couplings $d_V^g$ and $d_A^g$. The $\sum|\Delta {\cal A}|^2$ for correlation asymmetries (${\cal A}^{tW}$) are better
		than that of top polarization asymmetries (${\cal A}^t$)  for both couplings although not better than $W$ polarization asymmetries (${\cal A}^W$). However, the combined
	$|\Delta {\cal A}|^2$ of polarization and correlation asymmetries improve  significantly over the polarization asymmetries.

	\paragraph{Effect of $WWZ$ anomalous couplings in {\boldmath $u\bar{d} \to ZW^+$} :} 
	\begin{figure}[ht!]
		\centering
		\includegraphics[width=0.496\textwidth]{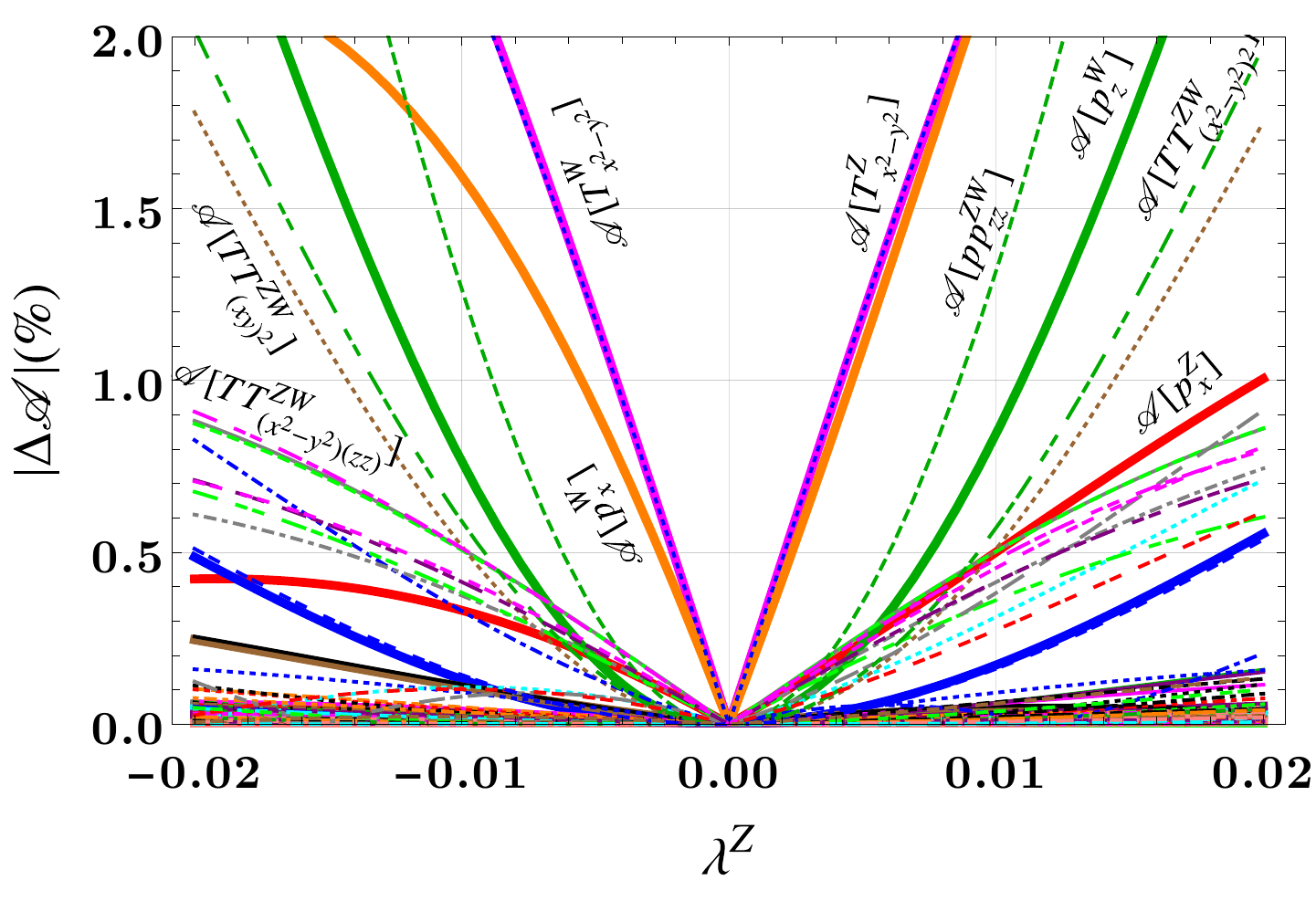}
		\includegraphics[width=0.496\textwidth]{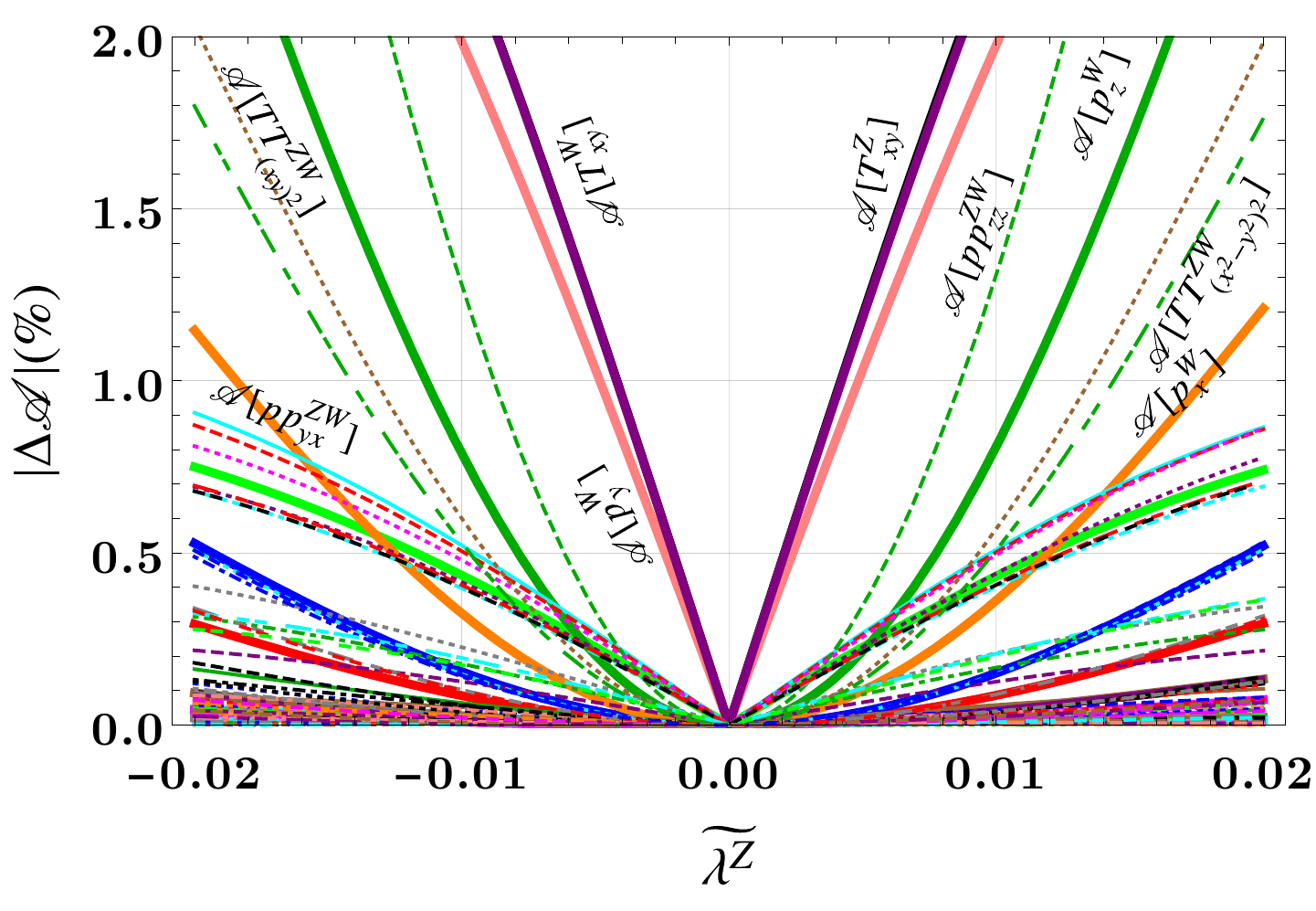}
		\includegraphics[width=0.496\textwidth]{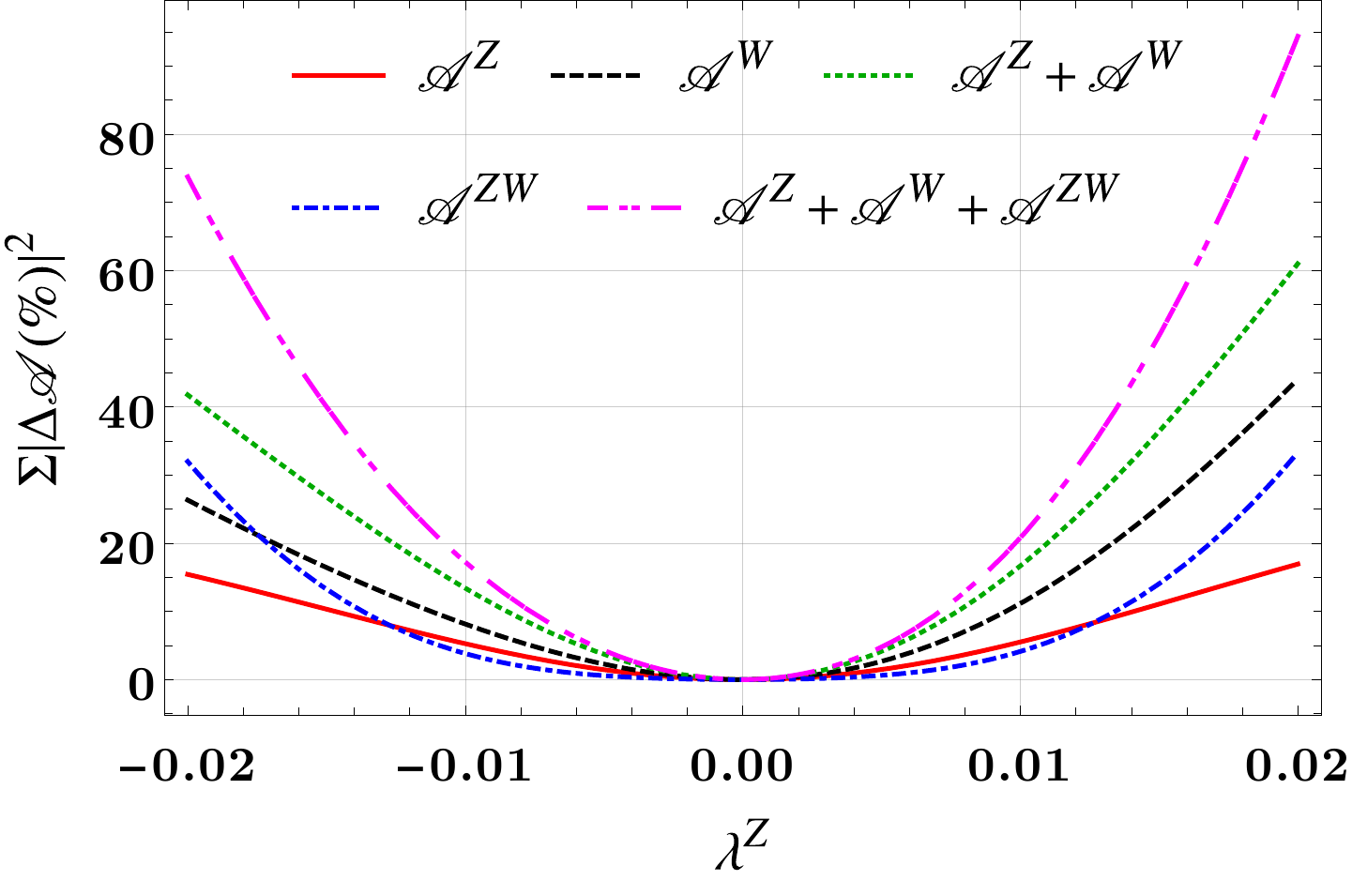}
		\includegraphics[width=0.496\textwidth]{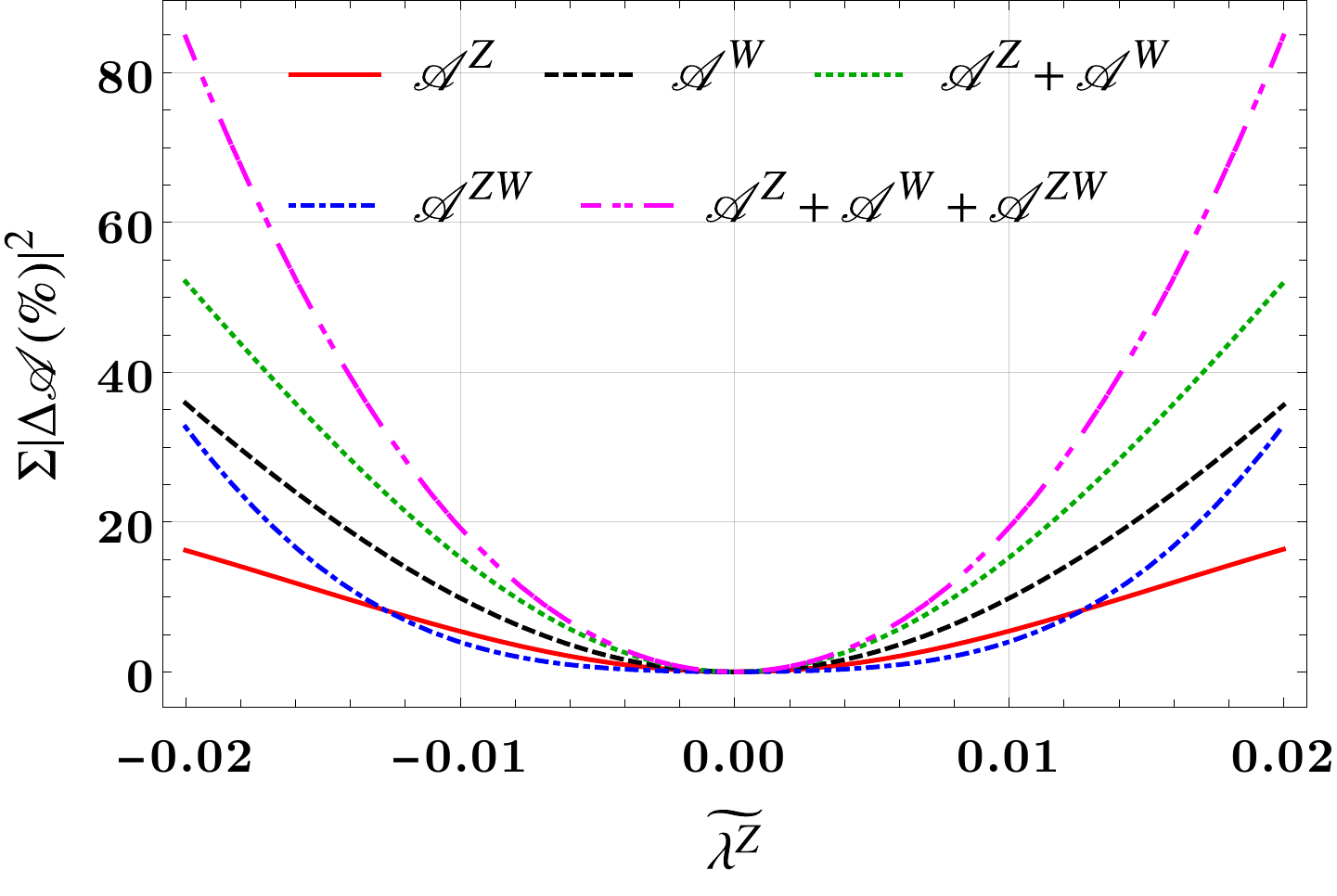}	
		\caption{\label{fig:dA-ZW} The absolute deviation of the asymmetry from the SM ($|\Delta {\cal A}|$) and their square sum ($\sum|\Delta {\cal A}|^2$) are shown as a function of anomalous couplings in the process  $u\bar{d}\to ZW^+$  at $\sqrt{\hat{s}}=1$ TeV.} 
	\end{figure}
	We choose the $u\bar{d} \to ZW^+$ partonic process as an example to see the effect of anomalous $WWZ$ couplings, one $CP$-even coupling $\lambda^Z$ and one $CP$-odd coupling $\wtil{\lambda^Z}$ (see Eq.~(\ref{eq:Lag-BSM})),   on the polarizations and spin correlations of $Z$ and $W^+$.  The $ZW^+$ production process including decay is chosen to be, 
	\begin{eqnarray}\label{eq:decay-process-bsm}
		u\bar{d} \to ZW^+&:&~~ Z\to e^+ e^-;~ W^+\to \mu^+ \nu_\mu .
	\end{eqnarray}
	In this process too, we use decay distribution of the daughters  of $Z$ and $W^+$ to obtain their polarization and correlation asymmetries by generating large number of events ($10^6$) in {\tt mg5\_aMC} at $\sqrt{\hat{s}}=1$ TeV with varying the anomalous couplings $\lambda^Z$ and $\wtil{\lambda^Z}$.
	Here also, all the asymmetries are entirely independent of each other, as the final states are entirely different.
	In Fig.~\ref{fig:dA-ZW}, we show the absolute deviation of all the polarization and correlation asymmetries  from the SM ($|\Delta {\cal A}|$) as a function of the couplings in {\em top-row}.
	We labeled the plot lines only for the asymmetries having $|\Delta {\cal A}| >1$
	within the range of couplings shown. 
	For the range of couplings shown in Fig.~\ref{fig:dA-ZW}, $|\Delta {\cal A}|$ for some parameters for $Z$ and $W^+$ happen to be same or very close, e.g., 
	$|\Delta {\cal A}[T_{x^2-y^2}^W]|\simeq |\Delta {\cal A}[T_{x^2-y^2}^Z]|$ appearing for $\lambda^Z$ in the {\em left-top-panel} and 
	$|\Delta {\cal A}[T_{xy}^W]|\simeq |\Delta {\cal A}[T_{xy}^Z]|$ appearing for $\wtil{\lambda^Z}$ in the {\em right-top-panel}.
	The correlation asymmetries also show large and comparable deviations as the polarization asymmetries do. 
	For example, $|\Delta {\cal A}|$ for  the correlation asymmetries 
	${\cal A}[pp_{zz}^{ZW}]$,   ${\cal A}[TT_{(xy)^2}^{ZW}]$, and   ${\cal A}[TT_{\l(x^2-y^2 \r)^2}^{ZW}]$ are large and comparable to  the polarization asymmetries ${\cal A}[p_x^W]$, ${\cal A}[p_z^W]$, and ${\cal A}[T_{x^2-y^2}^{W/Z}]$
	for small values of  $\lambda^Z$, see {\em left-top-panel} in Fig.~\ref{fig:dA-ZW}.
	Similar argument is true for $\wtil{\lambda^Z}$ shown in the {\em right-top-panel} of Fig.~\ref{fig:dA-ZW}. For example, 
	apart from the  polarization asymmetries ${\cal A}[p_y^W]$, ${\cal A}[p_z^W]$, and ${\cal A}[T_{xy}^{W/Z}]$,  the correlation asymmetries ${\cal A}[pp_{zz}^{ZW}]$,  ${\cal A}[TT_{(xy)^2}^{ZW}]$,  ${\cal A}[TT_{\l(x^2-y^2 \r)^2}^{ZW}]$, and ${\cal A}[pp_{yx}^{ZW}]$ also  show comparable deviation for small value of couplings, see {\em right-top-panel} in Fig.~\ref{fig:dA-ZW}.  
	The quadratic sum of $|\Delta {\cal A}|$ for the correlations asymmetries (${\cal A}^{ZW}$) is comparable to the polarization asymmetries of  (${\cal A}^Z$) and the polarization asymmetries of $W$ (${\cal A}^W$)  for both the couplings $\lambda^Z$ and $\wtil{\lambda^Z}$, see {\em bottom-panel} in Fig.~\ref{fig:dA-ZW}.
		The combinations of the correlations and the polarizations show significant
		improvement over the combined polarizations (${\cal A}^Z+{\cal A}^W$) for both the couplings $\lambda^Z$ and $\wtil{\lambda^Z}$.

	From the three cases we discussed above, it is evident that the spin correlations play a significant role along with the polarizations in probing new physics. The overall effect is more prominent  for $CP$-odd couplings ($d_A^\gamma$,  $d_A^g$ ,$\wtil{\lambda^Z}$) than the $CP$-even couplings ($d_V^\gamma$, $\lambda^Z$, $d_V^g$). 
	The partonic $ZW^+$ and $tW^-$ production processes are possible in a hadronic collider such as the LHC, where initial states are folded with parton distribution functions. In this case, the Lab frame will be different from the CM frame, and we have to consider the boost and rotations of the polarization and spin correlations through the relations given in section~\ref{sec:corrltn-in-labframe}. Realistic effects, 
	such as  initial state radiation (ISR), final state radiation (FSR), hadronization and detector effects, will affect the polarization and correlations.  All three processes contain missing neutrinos, which need to be reconstructed to obtain the rest frame of the particles whose polarization and correlations are to be obtained.   
	Reconstruction of missing neutrinos along with realistic effects would change the way $|\Delta A|$ depends on the anomalous couplings. Usually, these effects lead to lower sensitivity to anomalous couplings, but in some cases, the sensitivity can go up, e.g., see Ref.~\cite{Rahaman:2019lab} for $W$ polarizations. 
	
	\subsection{Note on higher order effect}\label{sec:higher-order-effect}
	Besides the realistic effects at colliders, higher order perturbative effects are important and should be taken into consideration while estimating the polarization and spin correlations in scattering processes~\cite{Rahaman:2018ujg,Rahaman:2019lab,Baglio:2018rcu}. Though the polarization and spin correlations receive corrections from higher order effect and changes from the tree level value, the formalism we discuss here holds in SM and BSM~\cite{Godbole:2006tq}.
Higher order radiative corrections to a full $2\to n$ process involve correction to the production part alone, the decay part alone, and the non-factorizable correction connecting the production and decay part. 

Corrections only to the production process  will change the polarization and spin correlation and hence change the decay product distributions. But the asymmetries from the decay product distribution do measure the polarization and spin correlation, including higher order effects. 

On the other hand, real emission from the decay products changes the energy and angle factorization of decay density matrices, hence changing the decay product distributions. However, whether QCD or QED, these corrections are very small and can be  neglected~\cite{,Jezabek:1988ja,Czarnecki:1990pe} if the decay products are color-neutral.

Non-factorizable higher order corrections are also negligible for off-shell weak boson~\cite{Beenakker:1997bp,Denner:1997ia} as well as top quark~\cite{Denner:1998rh,Macesanu:2001bj,Bernreuther:2004jv,Kolodziej:2005rx} production, while zero for on-shell production case
and will not affect the decay product distributions.

	\section{Summary}\label{sec:discussion}
	To summarize, we presented a formalism for the spin correlations of two-particle systems with  spins half-half, half-one, and one-one
	and showed the connection of these correlations with the joint angular distributions of the decay products by identifying the asymmetries for them. We validated the formalism in the SM in three processes, such as $e^-e^+\to t\bar{t}$, $gb\to tW^-$  and  $e^-e^+\to ZZ$ by calculating the correlations from the production process as well as from the decay angular distributions by generating events in {\tt mg5\_aMC} with complete decay chains with finite width effects of the mother particles. We find that although some individual polarizations vanish in the SM, their correlations do not, e.g., $CP$-odd polarizations 
	$p_y$ vanish for all particles, but $pp_{yy}$ do not vanish in any of the three scenarios. We then investigated the effect of some possible new physics on the correlations and polarizations  in three  processes, such as $\gamma t\bar{t}$ anomalous couplings in   $e^-e^+\to t\bar{t}$ process,  $gt\bar{t}$ anomalous couplings in $gb\to tW^-$ process,     and $W^+W^-Z$ anomalous couplings in $u\bar{d}\to ZW^+$ process. 
	We compare the polarization and correlation asymmetries individually in terms of absolute deviation from the SM ($|\Delta {\cal A}|$)  as well as combined way by taking the quadratic sum of $|\Delta {\cal A}|$ in all three processes. 
	The spin correlations have the potential to provide a significant improvement over the polarizations in probing the anomalous couplings. With a large set of spin correlations asymmetries, it will help to study a large number of anomalous couplings simultaneously in a given process. 
	It is straightforward to extend this method of describing spin correlations of two particles  to describe spin correlations of more than two particles~\cite{Rahaman:2022dwp}, which is beyond the scope of this paper.

	\vspace{0.5cm}
	\noindent \textbf{Acknowledgment:}
	\label{sec:ack}
	The work of RR is supported by funding available from the Department of Atomic Energy, Government of India, for the Regional Centre for Accelerator-based Particle Physics (RECAPP), Harish-Chandra Research Institute.
	The work of RKS is partially supported by SERB, DST, Government of India through the project EMR/2017/002778.  
	\appendix
	\section{Spin matrices and normalized decay density matrices}\label{sec:spin-matrices}
	The Pauli spin-$1/2$ matrices are
	\begin{equation}\label{eq:app:pauli-sigma}
		\tau_x=\begin{pmatrix} 0 & 1 \\ 1 & 0  \end{pmatrix},~~
		\tau_y=\begin{pmatrix} 0 & -i \\ i & 0  \end{pmatrix},~~
		\tau_z=\begin{pmatrix} 1 & 0 \\ 0 & -1  \end{pmatrix}.
	\end{equation}
	The spin-$1$ matrices are given by,
	\begin{eqnarray}
		S_x=\frac{1}{\sqrt{2}}
		\left(
		\begin{array}{ccc}
			0 & 1 & 0 \\
			1 & 0 & 1 \\
			0 & 1 & 0 \\
		\end{array}
		\right),~
		S_y=\frac{i}{\sqrt{2}}
		\left(
		\begin{array}{ccc}
			0 & -1 & 0 \\
			1 & 0 & -1 \\
			0 & 1 & 0 \\
		\end{array}
		\right)
		,~
		S_z=
		\left(
		\begin{array}{ccc}
			1 & 0 & 0 \\
			0 & 0 & 0 \\
			0 & 0 & -1 \\
		\end{array}
		\right).
	\end{eqnarray}

	The normalized decay density matrices at the helicity rest frame for  spin-$1/2$ particles and  spin-$1$ particles are given by~\cite{Boudjema:2009fz}, 
	\begin{equation}\label{eq:decay-half}
		\renewcommand{\arraystretch}{1.5}
		\Gamma_{(1)}(\lambda ,\lambda^\prime)=\left[
		\begin{tabular}{ll}
			$\frac{1+\alpha \cos\theta}{2}$ & $\frac{\alpha\sin\theta}{2} \ e^{i\phi}$\\
			$\frac{\alpha\sin\theta}{2} \ e^{-i\phi}$ & $\frac{1-\alpha \cos\theta}{2}$  \\
		\end{tabular} \right]~\text{and}
	\end{equation}
	\begin{equation}\label{eq:decay-one}
		\renewcommand{\arraystretch}{1.5}
		\Gamma_{(2)}(\lambda ,\lambda^\prime)=\left[
		\begin{tabular}{lll}
			$\frac{1+\delta+(1-3\delta)\cos^2\theta+2\alpha \cos\theta}{4}$ &
			$\frac{\sin\theta(\alpha+(1-3\delta)\cos\theta)}{2\sqrt{2}} \ e^{i\phi}$&
			$(1-3\delta)\frac{(1-\cos^2\theta)}{4} \ e^{i2\phi}$\\
			$\frac{\sin\theta(\alpha+(1-3\delta)\cos\theta)}{2\sqrt{2}} \ e^{-i\phi}$&
			$\delta+(1-3\delta)\frac{\sin^2\theta}{2}$ &
			$\frac{\sin\theta(\alpha-(1-3\delta)\cos\theta)}{2\sqrt{2}} \ e^{i\phi}$\\
			$(1-3\delta)\frac{(1-\cos^2\theta)}{4} \ e^{-i2\phi}$ &
			$\frac{\sin\theta(\alpha-(1-3\delta)\cos\theta)}{2\sqrt{2}} \ e^{-i\phi}$ &
			$\frac{1+\delta+(1-3\delta)\cos^2\theta-2\alpha\cos\theta}{4}$
		\end{tabular} \right], 
	\end{equation}
	respectively with $\theta$ and $\phi$ being the polar and azimuthal angle of the decay products, in the rest
	frame or helicity rest frame of the mother particles.
	The $\alpha$s are called spin analyzing power of the decay products. 
	For the spin-$1/2$ particle $f$ decaying to another spin-$1/2$ fermion $f^\prime$ and a spin-$1$ vector boson $V$ through the vertex structure $\bar f \gamma^\mu \ (C_L P_L
	+ C_R P_R) f^\prime V_\mu,~~ P_{L/R} = \frac{1}{2}\left(1\mp \gamma_5\right)$, the analyzing power $\alpha$ is given by~\cite{Boudjema:2009fz},
	\begin{equation}\label{eq:spin-half-alpha}
		\alpha_{\l(1\r)}=\frac{(C_R^2-C_L^2)(1-x_1^2-2x_2^2)\sqrt{1+(x_1^2-x_2^2)^2-2(x_1^2+x_2^2)
		}}{(C_R^2+C_L^2)(1-2x_1^2+x_2^2+x_1^2x_2^2+x_1^4-2x_2^4)-12C_LC_Rx_1x_2^2},
	\end{equation}
	where $x_i=m_i/m$ with $m_i$ as the mass of daughters and $m$ as the mass of mother particle. 
	In the case of  $t\to bW^+$ decay,  within the SM we have $\alpha \sim -0.396$. 
	For the case of $V\to ff^\prime$ decay through the same decay vertex as above, 
	the analyzing power $\alpha$ and $\delta$ are given by~\cite{Boudjema:2009fz},
	\begin{equation}
		\alpha_{(2)}=
		\frac{2(C_R^2-C_L^2)\sqrt{1+(x_1^2-x_2^2)^2-2(x_1^2+x_2^2)}}
		{12 C_LC_R x_1x_2+(C_R^2+C_L^2)[2-(x_1^2-x_2^2)^2+(x_1^2+x_2^2)]},
	\end{equation}
	\begin{equation}
		\delta=\frac{4C_LC_R x_1x_2
			+(C_R^2+C_L^2)[(x_1^2+x_2^2)-(x_1^2-x_2^2)^2]} {12 C_LC_R x_1x_2+(C_R^2+C_L^2)[2-(x_1^2-x_2^2)^2+(x_1^2+x_2^2)]}.
	\end{equation}
	%
	For massless final state fermions,
	$x_1\to0, \ x_2\to 0$; one obtains $\delta \to 0$ and 
	$\alpha \to (C_R^2-C_ L^2)/ (C_R^2+C_L^2)$. Furthermore, for the $W$ decay, within the SM we have $C_R=0$ and thus $\alpha=-1$. For the $Z$ decay to $l^+l^-$,  $\alpha \sim -0.2193$.
	
	\section{Spin correlations from production process in spin-$1/2$ -- spin-$1/2$ case}\label{app:corr-from-production}
	The harmitian spin-polarization density matrix for spin-$1/2$ -- spin-$1/2$ case in Eq.~(\ref{eq:spin-density-half-half}) will be expanded into
	{\footnotesize 
		\begin{eqnarray}\label{eq:pol-matrix-half-half}
			\renewcommand{\arraystretch}{1.8}
			&&P_{AB(1,1)}\l(\lambda_A,\lambda_A^\prime,\lambda_B,\lambda_B^\prime\r)= \nonumber\\
			&&	\dfrac{1}{4}\left[
			\begin{array}{llll}
				1+p_3^A+p_3^B+pp_{33}^{AB}  & ~~p_1^B+pp_{31}^{AB}-i (pp_{32}^{AB}+p_2^B) & ~~p_1^A+pp_{13}^{AB}-i (pp_{23}^{AB}+p_2^A) & pp_{11}^{AB}-pp_{22}^{AB}-i (pp_{12}^{AB}+pp_{21}^{AB}) \\
				& 1+p_3^A-p_3^B-pp_{33}^{AB} & pp_{11}^{AB}+pp_{22}^{AB}+i (pp_{12}^{AB}-pp_{21}^{AB}) & ~~p_1^A-pp_{13}^{AB}-i (p_2^A-pp_{23}^{AB}) \\
				&  & 1-p_3^A+p_3^B-pp_{33}^{AB} & p_1^B-pp_{31}^{AB} -i (p_2^B-pp_{32}^{AB}) \\
				&  & & 1-p_3^A-p_3^B+pp_{33}^{AB} \\
			\end{array}
			\right],\nonumber\\
		\end{eqnarray}
	}
	which can be compared to the normalized production density matrix ($\rho$)
	of $A$ and $B$  given below,
	\begin{equation}\label{eq:pol-matrix-half-half2}
		\renewcommand{\arraystretch}{1.5}
		P_{AB(1,1)}\l(\lambda_A,\lambda_A^\prime,\lambda_B,\lambda_B^\prime\r)=\dfrac{1}{\sigma_p}\left[
		\begin{array}{llll}
			\rho _{(++++)} & \rho _{(+++-)} & \rho _{(+-++)} & \rho _{(+-+-)} \\
			\rho _{(++-+)} & \rho _{(++--)} & \rho _{(+--+)} & \rho _{(+---)} \\
			\rho _{(-+++)} & \rho _{(-++-)} & \rho _{(--++)} & \rho _{(--+-)} \\
			\rho _{(-+-+)} & \rho _{(-+--)} & \rho _{(---+)} & \rho _{(----)} \\
		\end{array}
		\right].
	\end{equation}
	Here, we have used the notations $(+)\equiv (+1/2)$ and $(-)\equiv (-1/2)$ for the helicities of the spin-$1/2$
	particles $A$ and $B$.  
	One obtains the following relations for the polarizations and spin-correlations in terms of $\rho$:
	{\small 	
		\begin{eqnarray}\label{eq:pol-from-density-matriIx-half-half}
			p_{1}^A&=& \l[\rho _{(-+--)}+\rho _{(-+++)}+\rho _{(+---)}+\rho _{(+-++)}\r]/\sigma_p,
			\nonumber\\
			p_{2}^A&=& -i \l[\rho _{(-+--)}+\rho _{(-+++)}-\rho _{(+---)}-\rho _{(+-++)}\r]/\sigma_p,
			\nonumber\\
			p_{3}^A&=& \l[-\rho _{(----)}-\rho _{(--++)}+\rho _{(++--)}+\rho _{(++++)}\r]/\sigma_p,
			\nonumber\\
			p_{1}^B&=& \l[\rho _{(---+)}+\rho _{(--+-)}+\rho _{(++-+)}+\rho _{(+++-)}\r]/\sigma_p,
			\nonumber\\
			p_{2}^B&=& -i \l[\rho _{(---+)}-\rho _{(--+-)}+\rho _{(++-+)}-\rho _{(+++-)}\r]/\sigma_p,
			\nonumber\\
			p_{3}^B&=& \l[-\rho _{(----)}+\rho _{(--++)}-\rho _{(++--)}+\rho _{(++++)}\r]/\sigma_p,
			\nonumber\\
			pp_{11}^{AB}&=& \l[\rho _{(-+-+)}+\rho _{(-++-)}+\rho _{(+--+)}+\rho _{(+-+-)}\r]/\sigma_p,
			\nonumber\\
			pp_{12}^{AB}&=& -i \l[\rho _{(-+-+)}-\rho _{(-++-)}+\rho _{(+--+)}-\rho _{(+-+-)}\r]/\sigma_p,
			\nonumber\\
			pp_{13}^{AB}&=& \l[-\rho _{(-+--)}+\rho _{(-+++)}-\rho _{(+---)}+\rho _{(+-++)}\r]/\sigma_p,
			\nonumber\\
			pp_{21}^{AB}&=& -i \l[\rho _{(-+-+)}+\rho _{(-++-)}-\rho _{(+--+)}-\rho _{(+-+-)}\r]/\sigma_p,
			\nonumber\\
			pp_{22}^{AB}&=& \l[-\rho _{(-+-+)}+\rho _{(-++-)}+\rho _{(+--+)}-\rho _{(+-+-)}\r]/\sigma_p,
			\nonumber\\
			pp_{23}^{AB}&=& i \l[\rho _{(-+--)}-\rho _{(-+++)}-\rho _{(+---)}+\rho _{(+-++)}\r]/\sigma_p,
			\nonumber\\
			pp_{31}^{AB}&=& \l[-\rho _{(---+)}-\rho _{(--+-)}+\rho _{(++-+)}+\rho _{(+++-)}\r]/\sigma_p,
			\nonumber\\
			pp_{32}^{AB}&=& i \l[\rho _{(---+)}-\rho _{(--+-)}-\rho _{(++-+)}+\rho _{(+++-)}\r]/\sigma_p,
			\nonumber\\
			pp_{33}^{AB}&=& \l[\rho _{(----)}-\rho _{(--++)}-\rho _{(++--)}+\rho _{(++++)}\r]/\sigma_p .
		\end{eqnarray}
	}
	
	In case of spin-$1/2$ -- spin-$1$ and spin-$1$ -- spin$1$ correlations, one needs to use the trace equations 
	and  symmetry conditions (Eq.~(\ref{eq:constraints-half-one}), Eq.~(\ref{eq:constrain-TT-one-one})) along with Eq.~(\ref{eq:def-PAB}) to solve for the polarization and correlation parameters.

	\section{Asymmetries for tensor polarizations of spin-$1$ particles}\label{app:tensor-integrand-rule}
	The asymmetries for the five independent tensor ($T$) polarizations of a spin-$1$ particle $A$ are given by the following equations~\cite{Rahaman:2016pqj,Rahaman:2020jll},
	\begin{eqnarray}\label{eq:asym-T-12}
		{\cal A}[T_{12}^{A}] & \equiv & 
		\left( \int _{\theta_{a}=0}^{\pi }\int _{\phi_{a}=0}^{\frac{\pi }{2}}
		-\int _{\theta_{a}=0}^{\pi }\int _{\phi_{a}=\frac{\pi }{2}}^{\pi } \right. \nonumber\\
		&&\left. +\int _{\theta_{a}=0}^{\pi }\int _{\phi_{a}=\pi }^{\frac{3 \pi }{2}} 
		-\int _{\theta_{a}=0}^{\pi }\int _{\phi_{a}=\frac{3 \pi }{2}}^{2 \pi }\right)
		d\Omega_{a}	\l(\frac{1}{\sigma}\dfrac{d\sigma}{d\Omega_{a}}\r)  \nonumber\\
		&\equiv& \int_{a}^{T_{12}} d\Omega_{a}	\l(\frac{1}{\sigma}\dfrac{d\sigma}{d\Omega_{a}}\r)
		= \frac{2}{\pi } \sqrt{\frac{2}{3}} (1-3 \delta_{A} ) T_{12}^{A},\\
		{\cal A}[T_{13}^{A}] & \equiv &
		\left( \int _{\theta_{a}=0}^{\frac{\pi }{2}}\int _{\phi_{a}=-\frac{\pi }{2}}^{\frac{ \pi }{2}}
		-\int _{\theta_{a}=0}^{\frac{\pi }{2}}\int _{\phi_{a}=\frac{\pi }{2}}^{\frac{3\pi }{2}}\right. \nonumber\\
		&&\left.+\int _{\theta_{a}=\frac{\pi }{2}}^{\pi }\int _{\phi_{a}=\frac{\pi }{2}}^{\frac{3\pi }{2}}
		-\int _{\theta_{a}=\frac{\pi }{2}}^{\pi }\int _{\phi_{a}=-\frac{\pi }{2}}^{\frac{ \pi }{2}}\right)
		d\Omega_{a}	\l(\frac{1}{\sigma}\dfrac{d\sigma}{d\Omega_{a}}\r)  \nonumber\\
		&\equiv& \int_{a}^{T_{13}} d\Omega_{a}	\l(\frac{1}{\sigma}\dfrac{d\sigma}{d\Omega_{a}}\r)
		= \frac{2}{\pi } \sqrt{\frac{2}{3}} (1-3 \delta_{A} ) T_{13}^{A} ,\\
		{\cal A}[T_{23}^{A}] & \equiv & 
		\left(\int _{\theta_{a}=0}^{\frac{\pi }{2}}\int _{\phi_{a}=0}^{\pi }
		-\int _{\theta_{a}=0}^{\frac{\pi }{2}}\int _{\phi_{a}=\pi }^{2 \pi }\right. \nonumber\\
		&&\left.+\int _{\theta_{a}=\frac{\pi }{2}}^{\pi }\int _{\phi_{a}=\pi }^{2 \pi } 
		-\int _{\theta_{a}=\frac{\pi }{2}}^{\pi }\int _{\phi_{a}=0}^{\pi }\right)
		d\Omega_{a}	\l(\frac{1}{\sigma}\dfrac{d\sigma}{d\Omega_{a}}\r)  \nonumber\\
		&\equiv& \int_{a}^{T_{23}} d\Omega_{a}	\l(\frac{1}{\sigma}\dfrac{d\sigma}{d\Omega_{a}}\r)
		=\frac{2 }{\pi }\sqrt{\frac{2}{3}} (1-3 \delta _{A}) T_{23}^{A} ,\\
		{\cal A}[T_{11-22}^{A}] & \equiv & 
		\left( 
		\int _{\theta_{a}=0}^{\pi }\int _{\phi_{a}=-\frac{\pi }{4}}^{\frac{\pi }{4}}
		-\int _{\theta_{a}=0}^{\pi }\int _{\phi_{a}=\frac{\pi }{4}}^{\frac{3 \pi }{4}} \right. \nonumber\\
		&&\left. +\int _{\theta_{a}=0}^{\pi }\int _{\phi_{a}=\frac{3 \pi }{4}}^{\frac{5 \pi }{4}}
		-\int _{\theta_{a}=0}^{\pi }\int _{\phi_{a}=\frac{5 \pi }{4}}^{\frac{7 \pi }{4}} \right)
		d\Omega_{a}	\l(\frac{1}{\sigma}\dfrac{d\sigma}{d\Omega_{a}}\r)  \nonumber\\
		&\equiv& \int_{a}^{T_{11-22}}d\Omega_{a}	\l(\frac{1}{\sigma}\dfrac{d\sigma}{d\Omega_{a}}\r)
		=\frac{1}{\pi }\sqrt{\frac{2}{3}} (1-3 \delta _{A}) \left(T_{11-22}^{A}\right) ,\\
		{\cal A}[T_{33}^{A}] & \equiv & 
		\left(\int _{\theta_{a}=0}^{\frac{\pi }{3}}\int _{\phi_{a}=0}^{2 \pi }
		-\int _{\theta_{a}=\frac{\pi }{3}}^{\frac{2 \pi }{3}}\int _{\phi_{a}=0}^{2 \pi }
		+\int _{\theta_{a}=\frac{2 \pi }{3}}^{\pi }\int _{\phi_{a}=0}^{2 \pi }\right)
		d\Omega_{a}	\l(\frac{1}{\sigma}\dfrac{d\sigma}{d\Omega_{a}}\r)  \nonumber\\
		&\equiv& \int_{a}^{T_{33}} d\Omega_{a}	\l(\frac{1}{\sigma}\dfrac{d\sigma}{d\Omega_{a}}\r)
		=	\frac{3}{8}\sqrt{\frac{3}{2}} (1-3 \delta _{A}) T_{33}^{A} ,
		\label{eq:asym-T-33}
	\end{eqnarray}
	with $a$ being a daughter of the particle $A$.
	For the numerical purpose, these five asymmetries can also be obtained as,
	\begin{equation}\label{eq:numericA-tensor-one}
		{\cal A}_{m}[T^{A}]=\dfrac{\sigma\l({\cal C}_{m}^a >0\r)-\sigma\l({\cal C}_{m}^a <0\r)}
		{\sigma\l({\cal C}_{m}^a >0\r)+\sigma\l({\cal C}_{m}^a <0\r)}  ,~~m\in[1,2,3,4,5], 
	\end{equation}
	with
	\begin{equation}\label{eq:five-calC-app}
		{\cal C}_m  \in\l[c_xc_y,~c_xc_z,~c_yc_z,~c_x^2-c_y^2,~|\sqrt{c_x^2+c_y^2}|(4c_z^2-1)=\sin(3\theta) \r] .
	\end{equation}
	\bibliography{References}
	\bibliographystyle{utphysM}
\end{document}